\documentclass[a4paper,11pt]{article}
\pdfoutput=1 

\usepackage{jhepmod} 

\usepackage[T1]{fontenc} 

\usepackage{setspace}

\usepackage{subcaption}
\usepackage{physics}
\usepackage{dsfont}
\usepackage{tensor}
\usepackage[normalem]{ulem}
\usepackage{xcolor}
\usepackage{graphicx}
\usepackage[export]{adjustbox}
\usepackage{comment}
\usepackage{ifthen}
\usepackage[cal=cm]{mathalpha}
\usepackage{eulervm}
\usepackage{standalone}

\usepackage{tikz}
\usetikzlibrary{shapes.geometric}
\usetikzlibrary{decorations.markings}
\usetikzlibrary{decorations.pathmorphing}
\usetikzlibrary{calc}
\usetikzlibrary{3d}
\usetikzlibrary{intersections}
\usepackage{pgfplots}
\usetikzlibrary{pgfplots.fillbetween}
\tikzset{snake it/.style={decorate, decoration=snake}}
\usetikzlibrary{hobby}
\usepackage{tikz-3dplot}

\graphicspath{{./figs/}}

\colorlet{darkblue}{blue!70!black}
\colorlet{darkgreen}{green!50!black}
\colorlet{darkbrown}{brown!70!black}

\title{\boldmath A New Covariant Entropy Bound from Cauchy Slice Holography}

\author[a]{Ronak M Soni}
\author[a,b]{and Aron C. Wall}
\affiliation[a]{Department of Applied  Mathematics and Theoretical Physics, University of Cambridge,\\Wilberforce Road, Cambridge, CB3 0WA, United Kingdom}
\affiliation[b]{School of Natural Sciences, Institute for Advanced Study, Princeton, NJ 08540 USA.}

\emailAdd{ronakmsoni@gmail.com}
\emailAdd{aroncwall@gmail.com}

\abstract{
  We begin an investigation of a new holographic covariant entropy bound (HCEB) in gravity.
	This bound arises from Cauchy slice holography, a recently proposed duality between the bulk gravity theory and a `boundary' theory that lives on Cauchy slices.
    The HCEB is the logarithm of the maximum number of states of this theory that can pass through a given cut $\sigma$ of a Cauchy slice $\Sigma$ ($\sigma$ is thus a codimension-2 surface in the bulk).  We show that the bound depends only on the codimension-2 data on $\sigma$, and is thus independent of the choice of slice $\Sigma$.
	For classical states, the HCEB upper bounds the entanglement between two subregions of the boundary of $\Sigma$.
 
  We calculate the bound explicitly in pure three-dimensional GR with negative cosmological constant, where the Cauchy slice theory is the $T \overline{T}$-deformation of the dual CFT.
  We find that the imaginary energy eigenstates in the spectrum of the deformed theory play a crucial role for obtaining a valid bound in Lorentzian signature.
  Our bound agrees with the area of a surface at certain marginal and extremal surfaces, but differs elsewhere.
  In particular, it exceeds the area by an arbitrarily large amount for (anti)trapped surfaces, such as those that lie inside a black hole.
  Finally, we discuss how these results can be used to write down tensor networks corresponding to arbitrary Cauchy slices.
}

\begin{document}
\maketitle
\flushbottom

\section{Introduction}
One of the most important consequences of the discovery that black holes have entropy is the formulation of the holographic principle, see \cite{Bousso:2002ju} for a review.
This principle states, broadly, that in a theory of quantum gravity the information about a region of space should `live' on its boundary.
It was motivated originally by the fact that black hole entropy scales like the area of the horizon, and the covariant entropy bound conjectured as a result of this discovery \cite{Bousso:1999xy}, which suggests that the information about the interior is encoded locally on the event horizon --- $\mathcal{O} (1)$ qubits per Planck area.
This principle can be considered a constraint on UV completions of semi-classical gravity, and therefore it is interesting to improve it as far as possible.
In this work, we will advocate for a wrinkle in our understanding of the holographic principle: that the information may not be encoded on the boundary in an \emph{extensive} way, i.e.~there is no absolute limitation on the density of qubits per unit area.
Instead, bounds of the form $S \le A/4 G_{N}$ will hold only for certain classes of codimension 2 surfaces.  We will argue for this new wrinkle by showing that Cauchy slice holography \cite{Araujo-Regado:2022gvw,Araujo-Regado:2022jpj,Khan:2023ljg} implies an improved entropy bound, that can disagree with the area by an arbitrarily large amount.  This falsifies the hypothesis of \cite{Bianchi:2012ev}, that entanglement is proportional to area for general regions of quantum gravity.

The clearest version of the holographic principle is the AdS/CFT conjecture.
While the AdS/CFT conjecture has the advantage of being precise, checkable and possible to derive from string theory, it is also much less general than the holographic principle itself since it only applies to the asymptotic boundary of a particular class of spacetimes.
Significant progress on this was made by the discovery of the $T \overline{T}$ deformation \cite{Zamolodchikov:2004ce,Lechner:2006kb,Freidel:2008sh,Dubovsky:2012wk,Smirnov:2016lqw,Cavaglia:2016oda,Dubovsky:2017cnj,Cardy:2018sdv,Dubovsky:2018bmo,Gorbenko:2018oov,McGough:2016lol,Kraus:2018xrn} --- see \cite{Jiang:2019epa} for a review --- which is an integrable deformation of two-dimensional QFTs that results in a field theory that is not the relevant deformation of some CFT.
Considered rigorously, this $T \overline{T}$ deformation is defined nonperturbatively only on flat manifolds for two-dimensional theories.
However, for a theory with a large number of degrees of freedom, it can also be defined  perturbatively in a $1/N$ (or $1/c$ in two dimensions) expansion in arbitrary dimensions and on arbitrary manifolds, see especially \cite{Hartman:2018tkw,Taylor:2018xcy}.
In this limit, it was noticed that the deformed theory is dual to the bulk theory with a cut-off surface at finite distance \cite{McGough:2016lol,Shyam:2017znq,Shyam:2017qlr,Kraus:2018xrn}.
This can also be extended to other asymptotics for the bulk, see \cite{Gorbenko:2018oov,Lewkowycz:2019xse,Shyam:2021ciy,Coleman:2021nor,Batra:2024kjl} for the case of positive cosmological constant and \cite{Coleman:2022lii} for the zero cosmological constant case.

Thus, the $T \overline{T}$ deformation gives an example of the holographic principle that goes far beyond AdS/CFT proper.
Further, it is capable of exploring physics well below the AdS or dS scale, breaking down only near the Planck or string scale.
Unlike AdS/CFT, however, the boundary theory can currently only be defined in perturbation theory.\footnote{
	See \cite{Batra:2024kjl} for a version that is guaranteed to be well-defined in the UV, even though the details of the theory need to be fine-tuned in perturbation theory.
}
If the deformed theory were successfully UV-completed, it would give a non-perturbative definition of bulk quantum gravity \cite{Araujo-Regado:2022gvw}.

The dual theory is able to reproduce the thermodynamics of the bulk spacetime.
This is possible due to the fact that the deformation is a form of spectral flow --- i.e. different energy levels don't interact as we deform the theory --- and so it is possible to calculate the density of states in the deformed theory using the spectrum of the undeformed theory.
An important feature here is that this procedure does not require equating entropy to area \cite{McGough:2016lol,Caputa:2020fbc}, as we review in section \ref{ssec:entropy-cyl} below.\footnote{
	In fact, in the case of positive cosmological constant, it was found, under some assumptions, that the entropy exceeded area quite drastically \cite{Shyam:2021ciy,Coleman:2021nor,Batra:2024kjl}.
}
In this paper, we argue that this is a `feature rather than a bug,' and show that it implies a refined entropy bound that differs from the area except at (certain classes of) marginal and extremal surfaces.

For this, we use the fact that the deformed theory can also live on a Cauchy slice \cite{Caputa:2020fbc,Araujo-Regado:2022gvw}.
In this case, the partition function of the deformed theory can be related to the Wheeler-DeWitt (WdW) wavefunction of a state \cite{Freidel:2008sh,Araujo-Regado:2022gvw,Witten:2022xxp}.
For a CFT state $\ket{\Psi}$ prepared by a Euclidean path integral and a slice $\Sigma$ with metric $g$, the inner product is
\begin{equation}
  \braket{g}{\Psi} = \quad
		\begin{tikzpicture}[baseline={(0,-.5)}]
		\draw[thick,fill=gray,fill opacity=.5] (0,0) arc (180:360:1.5) coordinate[pos=.5] (psi) to[out=180,in=-15] coordinate[above] (g) (1.5,.3) to[out=165,in=0] (1,.4) to[out=180,in=15] (.5,.25) to[out=195,in=0] (0,0);
		\node[label=below:$\Psi$] at (psi) {};
		\node[label=above:$g$] at (g) {};
		\draw[->] (3.5,-1) -- node[pos=0,below] {\tiny ket} (3.5,-.5);
		\draw[->] (3.5,.6) -- node[pos=0,above] {\tiny bra} (3.5,0.1);
		\node (dyn) at (1.5,-.65) {\tiny on-shell geometry};
	\end{tikzpicture}
	\quad = \quad 
		\begin{tikzpicture}[baseline={(0,-.5)}]
		\draw[thin,opacity=.5,gray] (0,0) arc (180:360:1.5) node[pos=.5,below] {$\Psi$};
		\draw[thick] (3,0) to[out=180,in=-15] node[above,text=black] {$g$} (1.5,.3) to[out=165,in=0] (1,.4) to[out=180,in=15] (.5,.25) to[out=195,in=0] (0,0);

		\draw[->] (1.5,1) -- node[pos=0,left] {\tiny path integral} (2,1);

	\end{tikzpicture}
	\quad = \tr \left( F_{\Psi} \mathds{T} [g] \right),
  \label{eqn:csh}
\end{equation}
where $\mathds{T}[g]$ is the transition matrix of the theory on the Cauchy slice and $F_{\Psi}$ is an operator whose matrix elements are related to the wavefunction of $\ket{\Psi}$.

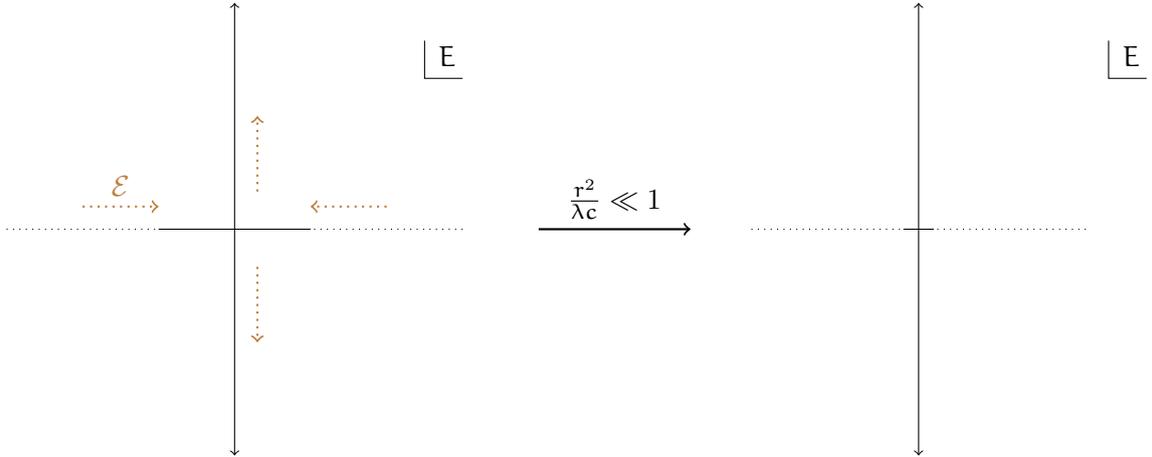
\begin{figure}[h!]
  \centering
	  \begin{tikzpicture}
    \draw (-1,0) -- (1,0);
    \draw[->] (0,0) -- (0, 3);
    \draw[->] (0,0) -- (0,-3);
    \draw[dotted] (-3,0) -- (-1,0);
    \draw[dotted] ( 3,0) -- ( 1,0);
    \draw (2.5,2.5) -- (2.5,2) -- (3,2);
    \node (E) at (2.8,2.3) {$E$};
    \draw[dotted,thick,brown,->] (-2,.3) -- node[pos=.5,above] {$\mathcal{E}$} (-1,.3);
    \draw[dotted,thick,brown,->] ( 2,.3) -- ( 1,.3);
    \draw[dotted,thick,brown,->] (.3, .5) -- (.3, 1.5);
    \draw[dotted,thick,brown,->] (.3,-.5) -- (.3,-1.5);
    
    \draw[thick,->] (4,0) -- node[pos=.5,above] {$\frac{r^{2}}{\lambda c} \ll 1$} (6,0);

    \begin{scope}[shift={(9,0)}]
    \draw (-.2,0) -- (.2,0);
    \draw[->] (0,0) -- (0, 3);
    \draw[->] (0,0) -- (0,-3);
    \draw[dotted] (-2.2,0) -- (-.2,0);
    \draw[dotted] ( 2.2,0) -- ( .2,0);
    \draw (2.5,2.5) -- (2.5,2) -- (3,2);
    \node (E) at (2.8,2.3) {$E$};
    \end{scope}
  \end{tikzpicture}
  \caption{The spectrum of the deformed CFT on a cylinder, at zero angular momentum: $J = 0$. Brown dotted lines denote the direction of increasing $\mathcal{E}$, the energy level of the seed CFT that a given state flows to as $\lambda/r^2 \to 0$. Black dotted lines indicate a sparse spectrum and solid lines a non-sparse spectrum with density of states given by the Cardy formula \eqref{eqn:cardy-dos}. As we take $r$ small, the beginning of the dense part of the spectrum tends to $E = 0$.}
  \label{fig:cross}
\end{figure}

The one-point function of the stress tensor of the deformed theory is the Brown-York stress tensor, or equivalently the `momentum' conjugate to the metric in the ADM formulation, which is given by the extrinsic curvature of $\Sigma$ in the on-shell geometry above.
This extrinsic curvature is with respect to the outward normal, which can be either spacelike or timelike, depending on the signature of the on-shell geometry in the neighbourhood of $\Sigma$.
As a result, the energy eigenvalues of the deformed theory can be either real (for a spacelike normal) or imaginary (for a timelike normal).
These imaginary eigenvalues are a well-known feature of the $T \overline{T}$-deformed theory, and Cauchy slice holography finds a natural role for them.
The full spectrum of the deformed theory is shown in figure \ref{fig:cross}.

Apart from having both real and energy energies, this spectrum also exhibits a $\mathbb{Z}_2 \times \mathbb{Z}_2$ discrete symmetry group generated by reflections of $E$ about the real and imaginary axis respectively.  These symmetries are:
\begin{enumerate}
\item an $E \to E^*$ symmetry, due to CPT symmetry,
\item an $E \to -E^*$ symmetry, which on a Cauchy slice $\Sigma$ relates to the reality conditions of bulk fields in a unitary bulk theory, and 
\item the product of the above is a $E \to -E$ symmetry: this corresponds to the fact that flipping the direction of the outward normal (something we're allowed to consider) also changes the sign of the extrinsic curvature.
\end{enumerate}
In the real part of the spectrum, all but the first symmetry is spontaneously broken, while in the imaginary part all but the second is spontaneously broken.

\begin{figure}[h!]
  \centering
	\begin{subfigure}[t]{.4\textwidth}
		\centering
		\includegraphics{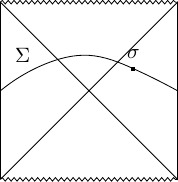}
		\caption{The entropy bound is the maximum number of states passing through a Cauchy slice $\Sigma \supset \sigma$ between the two asymptotic boundaries.}
		\label{fig:Ss-eg}
	\end{subfigure}
	\quad
	\begin{subfigure}[t]{.55\textwidth}
				\begin{tikzpicture}[scale=.75]
			\draw (4,0) -- (4,4);
			\draw (1.26,1.26) -- (4,4);
			\draw (1.26,2.74) -- (4,0);
			\draw[decorate,decoration={zigzag,amplitude=.3mm,segment length=.5mm}] (2,0) -- (4,0);
			\draw[decorate,decoration={zigzag,amplitude=.3mm,segment length=.5mm}] (2,4) -- (4,4);
			\draw[red,ultra thick] (2,0) to[out=135,in=-135] (2,4);

			\node[fill, inner sep=1,label=above:$\sigma$] (s1) at (3,2.5) {};

			\draw[->] (5,2) -- (6,2);

			\begin{scope}[xshift=7cm,scale=2]
			  \draw (0,0) -- (0,2) -- (2,0) -- (2,2) -- (0,0);
			  \draw[decorate,decoration={snake,amplitude=.3mm,segment length=1mm}] (0,2) -- (2,2);
			  \draw[decorate,decoration={snake,amplitude=.3mm,segment length=1mm}] (0,0) -- (2,0);

				\node[fill, inner sep=1,label=above:$\sigma$] (s1) at (1.5,1.25) {};
			\end{scope}
		\end{tikzpicture}
		\caption{The entropy bound can also be thought of as the maximum entropy of a spacetime in which $\sigma$ can be embedded. The left spacetime is a pure-state black hole with an end-of-the-world brane, and has entropy $0$, whereas the right is a two-sided black hole with a large entropy. They are the same near $\sigma$.}
		\label{fig:bulk-bound}
	\end{subfigure}
	\caption{The two ways of thinking about our entropy bound. Left: the `boundary' perspective. Right: the `bulk' perspective.}
\end{figure}

At any given value of $E$ and $J$, there are a finite number of possible states (at least in the pure gravity $T \overline{T}$ spectrum).  For sufficiently classical boundary states that are dual to a single bulk geometry, this picture suggests a natural algorithm to assign an entropy to an arbitrary codimension-two slice $\sigma$ in the bulk.
Consider a Cauchy slice $\Sigma$ which is a Euclidean manifold with two boundaries; and let $\sigma \in \Sigma$ be homologous to either of these boundaries, as shown in figure \ref{fig:Ss-eg}.
Then the path integral of the $T \overline{T}$ theory on $\Sigma$ can be interpreted as a transition matrix $\mathds{T}[\Sigma]$ between the two boundaries, and the entropy we assign to $\sigma$ is the logarithm of the maximum rank of $\mathds{T}[\Sigma]$ consistent with the data on $\sigma$.
(This algorithm is closely related to the microcanonical entropy, which counts the maximum number of states compatible with given macrostate parameters, e.g.~energy, angular momentum etc., as we review in section \ref{ssec:cft-bd}).

There is a second algorithm for associating an entropy with a surface $\sigma$, from a purely bulk perspective.
Fix the data on the codimension-two surface $\sigma$, and consider all possible on-shell spacetimes $\mathcal{M} \supset \sigma$ compatible with this data.
An entropy bound we can associate to $\sigma$ is the maximum fine-grained (HRT) entropy among all such $\mathcal{M}$.
An example is shown in figure \ref{fig:bulk-bound}.
This question is similar to the ``outer entropy'' explored in \cite{Engelhardt:2018kcs,Nomura:2018aus,Bousso:2018fou,Bousso:2019dxk,Wang:2020vxc}, but unlike these works we fix data only on a codimension-two surface, not in an entire wedge.
We especially note \cite{Bousso:2018fou}, which calculated the outer entropy for general surfaces.
Our explicit bounds agree with theirs where they are comparable, but our formalism applies to a larger class of surfaces in principle.

We will argue in section \ref{sec:bound-gen} that these two (logically distinct) algorithms should always give the same answer, as a consequence of holography.

We are able to calculate this entropy bound, which we call the \emph{holographic covariant entropy bound} (HCEB).
In the special case of a rotationally symmetric codimension-2 surface $\sigma$ in solutions of pure 3d GR, we find a completely explicit formula.
The quasi-local data on $\sigma$ is the area $A = 2\pi r$ (really a length), the null expansions $\theta_{\pm}$ and the twist $\chi$.
In terms of these and the AdS radius $\ell$, the formula is
	\begin{align}
	  S_{\mathrm{bd}} (\sigma) &= 2\pi \sqrt{\frac{\pi \ell}{2 G_{N}} \left( \mathcal{E} + \sqrt{\mathcal{E}^{2} - J^{2}} \right)}, \nonumber\\
														\text{where:} \qquad & 16 \pi G_{N} \ell \mathcal{E} \equiv r^{2} \left( 1 - 2 \ell^{2} g^{+-} \theta_{+} \theta_{-} \right) + \frac{\ell^{2} \chi^{2}}{4}, \nonumber\\
														 & 16 \pi G_{N} J = - r \chi.
	  \label{eqn:S-sigma-intro}
	\end{align}
In the above expressions, $\mathcal{E},J$ are (up to factors) the ADM mass and angular momentum of the spacetime.
This formula simplifies in the case without angular momentum:
	\begin{equation}
		S_{\mathrm{bd}} (\sigma) = \frac{\sqrt{A^{2} (1 - 2 g^{+-} \theta_{+} \theta_{-})}}{4 G_{N}},
		\label{eqn:S-sigma-2}
	\end{equation}
and it also simplifies in the case where either $\theta_{+}$ or $\theta_{-}$ vanishes, i.e.~when $\sigma$ is either marginal or extremal:
\begin{equation}\label{marginal}
S_{\mathrm{bd}} (\sigma) \,=\, \frac{1}{4G_N} \max(A, \ell |\chi|/2),
\end{equation}

To lift rotational symmetry, we then use the fact that classical 3d GR can be rewritten as a Chern-Simons theory: the value of $\mathcal{E},J$ can be extracted from the value of the Wilson loop around $\sigma$, and the entropy bound is still the first line of \eqref{eqn:S-sigma-intro}.
Because the Wilson loop is a path-ordered integral of an exponential, the bound depends on the data on $\sigma$ in a non-local way.

This is quite a bit more complicated than $S = A/4 G_{N}$.
However, the complexity of our bound is not evidence against it, because the bound is a derived output from our theoretical assumptions, not a conjectural input.
This is very different from the motivation of many previous entropy bounds, where the goal was to construct the simplest hypothesis not ruled out by a suite of thought experiments \cite{Bousso:2002ju}.
Further, for a marginally trapped $\sigma$ satisfying a certain inequality, our bound exactly agrees with the usual Bekenstein-Hawking formula:
\begin{equation}\label{Aover4}
	S_{\mathrm{bd}} = A/4 G_{N}.
\end{equation}
The familiar formula is a special case of our bound.

More generally, there are modifications to \eqref{Aover4} even for some marginal surfaces $\sigma$, due to the alternative to the area in \eqref{marginal} for surfaces with sufficiently high twist $\chi$.  As a result, the area bound works in ``subcritical'' cases where $\sigma$ has a sufficiently low angular momentum ($J < r^2$ in certain units), such that it is possible to think of $\sigma$ as lying on the ``outer horizon'' of a BTZ black hole.
However, in ``supercritical'' cases (where $|J| > r^{2}$), $\sigma$ has the right data to lie on the ``inner horizon'' of a BTZ black hole, and in this case we find a modified entropy formula
\begin{equation}
S_{\mathrm{bd}} = \overline{A}/4 G_{N}.
\end{equation}
defined in terms of a ``dual area'': $\overline{A} \propto |J|/A$, which is larger than $A$ iff $\sigma$ is supercritical.
In other words, the entropy of an ``inner horizon'' is proportional to the area of the (sometimes much larger) outer horizon.\footnote{Although we only consider pure gravity in this paper, we expect that a similar effect would apply to surfaces with sufficiently large electric charge, so as to be inner horizons of a Reissner–Nordstr\"{o}m black hole.}

Suppose we restrict attention to the subcritical case, and consider surfaces $\sigma$ for which the extrinsic curvature is not null.
For ``normal'' surfaces (those whose extrinsic curvature is spacelike), we find a weaker bound.
In other words, such surfaces always satisfy 
\begin{equation}
S_{\mathrm{bd}} < A/4 G_{N},
\end{equation}
so that, although a naive entropy bound is satisfied, it is not the tightest possible bound that can be written down.
On the other hand (anti)trapped surfaces, with timelike extrinsic curvature, can actually exceed the Bekenstein-Hawking entropy density:
\begin{equation}\label{exceeds}
S_{\mathrm{bd}} > A/4 G_{N}\,!
\end{equation}
Although this seems shocking, we will argue that such violations are required to develop a holographic picture of the black hole interior.
We will see that in order to obtain \eqref{exceeds} for
trapped surfaces, it is necessary to consider microcanonical ensembles of states with imaginary energy, as we have already indicated above.

Finally, we build 1-dimensional tensor networks based on our results, i.e.~we re-write the $T \overline{T}$ partition function as a product of matrices with variable bond dimension.
The basic idea is to use \eqref{eqn:S-sigma-intro} as the bond dimension of the bond associated to $\sigma$.\footnote{
	A closely related construction appeared in \cite{Chandra:2023dgq}.
	The authors of that work extracted $\mathcal{E},J$ in \eqref{eqn:S-sigma-intro} from the CFT stress tensor expectation values in the Euclidean past, and created tensor networks with the same bond dimensions as ours, but only for time-reversal symmetric slices.  For some previous approaches to building holographic tensor networks (and other related models) see \cite{Swingle:2009bg,Swingle:2012wq,Almheiri:2014lwa,Pastawski:2015qua,Miyaji:2015yva,Hayden:2016cfa,Cotler:2017erl,Kohler:2018kqk,Bao:2018pvs,Bao:2019fpq,Caputa:2017urj,Jafari:2019qns,Akers:2024ixq,Chen:2024unp,Akers:2024wab}.
}
Unlike most previous holographic tensor networks, these are not restricted to time-reversal symmetric Cauchy slices.\footnote{
	Recently, the tensor networks of \cite{Chen:2024unp,Akers:2024wab} have been proposed which correspond not to any particular Cauchy slice but to an entire Wheeler-DeWitt patch.
	Our tensor network, on the other hand, does correspond to particular Cauchy slices.
}
In particular, in the case of a black hole in the classical limit, we can get arbitrarily close to the singularity.
In this limit the area becomes arbitrarily small, and yet the minimal required bond dimension (which is still the boundary entropy $S_\mathrm{CFT} = A_\mathrm{HRT}/4G_N$) remains the same!  Without the non-extensive property for trapped surfaces \eqref{exceeds}, it would be impossible to construct a tensor network that goes into the black hole region.
As a corollary, without \eqref{exceeds} it would also not be possible for such a tensor network to capture the full dynamical evolution of general relativity, or in other words it cannot satisfy the Hamiltonian constraint equation.
But the $T \overline{T}$ partition function does satisfy the GR constraint equations; it compensates for this by having an entropy that is not always extensive.

The plan of the paper is as follows:
We introduce the $T \overline{T}$ deformation in section \ref{sec:TTbar}.
Then, in section \ref{sec:bound-gen}, we define the HCEB.
In section \ref{sec:ttb-spect}, we put forward our proposal for the spectrum of the deformed theory and then work out explicit formulas for pure three-dimensional GR in section \ref{sec:ent-bd}.
We use these results to write down tensor networks in section \ref{sec:btz-tn} and discuss simple extensions in section \ref{sec:extensions}.
We end with discussion in section \ref{sec:disc}.

Notation, conventions and standard formulae can be found in appendix \ref{app:convs}.
In appendix \ref{app:spectrum} we explain how the complex $T \overline{T}$ spectrum can be understood as self-adjoint with respect to an indefinite norm.
Finally, appendix \ref{app:eq-2} provides a second argument for the equivalence of the bulk and boundary algorithms for the entropy bound.

In a companion paper \cite{Soni:2024}, we will extend our 3D gravity results to situations involving matter fields satisfying the (quantum) dominant energy condition.
The existence of local degrees of freedom in the bulk modifies the story in two key ways:  first, it helps to clarify that the $S_\mathrm{bd}$ derived in this paper is a true \emph{coarse-grained} entropy, which can dynamically evolve across matter pulses, and become greater than $S_\mathrm{HRT}$.
Secondly, while the pure gravity $S_\mathrm{bd}$ derived in this paper continues to hold for marginal and normal surfaces (and indeed, can sometimes be tightened by a small amount) in the case of trapped surfaces, there is a very dramatic change causing the bound to diverge all the way to $S_\mathrm{bd} = +\infty$.

\section{$T \overline{T}$ Deformations and Cauchy Slice Holography} \label{sec:TTbar}
The $T \overline{T}$ deformation is a way of flowing ``up'' an RG flow instead of ``down.''
It is defined by the following irrelevant deformation:\footnote{There are many conventions for the stress tensor as well as for the normalisation of $\lambda$.
We follow the conventions of \cite{Torroba:2022jrk} with $\alpha = 1$. Our stress tensor conventions are listed in \eqref{eqn:T-conv}.}
\begin{equation}
  \partial_{\lambda} \log Z_{\lambda} = \int \sqrt{g} \left\{ \left\langle \det \mathcal{T}_{\lambda} \right\rangle_{\lambda} - \frac{c}{48\pi\lambda} R \right\} \qquad \det \mathcal{T} \equiv \frac{1}{2} \epsilon^{\mu\alpha} \epsilon^{\nu\beta} \mathcal{T}_{\mu\nu} \mathcal{T}_{\alpha\beta}
  \label{eqn:ttbar-eqn}
\end{equation}
along with the boundary condition that at $\lambda = 0$ we recover the partition function of some seed QFT.
We will take this seed QFT to be a holographic CFT dual to pure gravity at low energies except when specified otherwise.
Here, the RHS involves the expectation value of the stress tensor of the deformed theory \emph{in} the deformed theory.

This deformation results in a well-defined theory for arbitrary QFTs in flat space \cite{Zamolodchikov:2004ce,Smirnov:2016lqw,Cavaglia:2016oda,Cardy:2018sdv}.
A deformed theory can also be defined perturbatively in the $1/c$ expansion, with $c$ the central charge of the seed CFT, on curved spaces \cite{McGough:2016lol,Kraus:2018xrn}.
Here, we must take an order of limits where $c \to \infty, \lambda \to 0, \lambda c$ finite.
A similar expansion allows one to define it in higher dimensions \cite{Taylor:2018xcy,Hartman:2018tkw,Araujo-Regado:2022gvw}.
The main obstacle in defining the theory is arguing that the $T \overline{T}$ operator (the $\det \mathcal{T}$) can be defined in the deformed theory, which need not be a QFT.
In flat two-dimensional space, one can show that all possible theory-dependent contact terms are total derivatives, allowing for a universal definition.
But this is not the case in curved spaces \cite{Jiang:2019tcq}, meaning that one has to define the operator separately for every seed QFT and (in general) every value of $\lambda$.\footnote{
	This definition might not be continuous across phase transitions in $\lambda$-space.
}
In the large-$c$ limit, these problems are subleading in the $1/c$ expansion for any geometry \cite{Taylor:2018xcy,Hartman:2018tkw,Araujo-Regado:2022gvw}.
\cite{Batra:2024kjl} argued that, independent of these details, there is a well-defined deformation.

A particularly useful way to recast the flow equation \eqref{eqn:ttbar-eqn} in the case when the seed theory is a CFT is as
\begin{equation}
  \langle \mathcal{T}^{\mu}_{\mu} \rangle - 2 \lambda \langle \det \mathcal{T} \rangle + \frac{c}{24\pi} R = 0.
  \label{eqn:dim-anal-eqn}
\end{equation}
This is known as the trace-flow equation.
An integrated version of this equation is true by dimensional analysis because $\lambda$ is the only scale in a deformed CFT; it turns out to be true at large $c$ also at every point \cite{McGough:2016lol,Kraus:2018xrn}.
It will be particularly useful for us to redefine the stress tensor
\begin{equation}
  \mathcal{T}_{\mu\nu} = T_{\mu\nu} + \frac{1}{2\lambda} g_{\mu\nu}
  \label{eqn:T-shift}
\end{equation}
by adding a cosmological constant counterterm to the partition function.
This removes the linear term in the trace-flow equation,
\begin{equation}
  2 \lambda \langle \det T \rangle - \frac{c}{24\pi} R - \frac{1}{2\lambda} = 0.
  \label{eqn:tr-flow-eqn}
\end{equation}
We will henceforth drop expectation value brackets and call \eqref{eqn:tr-flow-eqn} the trace-flow equation, despite the fact that we have got rid of the eponymous trace.

Much of the interest in these deformed theories comes from the fact that they are integrable, in the sense that S-matrices and energy levels of the deformed theories are related in a simple way to those of the undeformed theory.
For a seed CFT, the deformed energy levels on a cylinder of radius $r$ are\footnote{If one interprets the radical sign in \eqref{cross} as a bivalued map from $\mathds{C}$ to a Riemann surface, then the $\pm$ symbol would be redundant. However, in this paper, we write $\pm$ whenever both signs of the square-root can be taken, and use the radical sign alone for just the positive root (which implicitly requires the argument to be non-negative).}
\begin{equation}
  E_{\lambda} (r,\mathcal{E},J) = \pm \frac{\pi}{\sqrt{\lambda}} \sqrt{\frac{r^{2}}{\lambda} - 4 \mathcal{E} + 4 \frac{\lambda}{r^{2}} J^{2}}, \qquad \mathcal{E} \equiv \frac{h + \bar{h} - \frac{c}{12}}{2\pi},\ J \equiv \frac{h - \bar{h}}{2\pi}.
  \label{eqn:deformed-energy}
\end{equation}
Here, $h,\bar{h}$ are the conformal dimensions of the CFT state.
The $-$ branch of the square root limits to the CFT energy at $\lambda \to 0$ after reversing the shift in \eqref{eqn:T-shift}.
The $+$ branch diverges in the same limit.

As one increases the dimensionless parameter $\lambda/r^{2}$ from $0$, the energies are real, then they become imaginary, and then they become real again.
The first set of real energies satisfies $|J| < r^{2}/2\lambda$; we will refer to these as \emph{subcritical}.
The second set satisfies $|J| > r^{2}/2\lambda$; we will refer to these as \emph{supercritical}.
For later convenience, let us note that this energy spectrum has an unexpected $\mathds{Z}_{2}$ symmetry between the subcritical and supercritical cases under $r \to 2 \lambda |J|/r$, i.e. $r^{2}/\lambda \to 4 \lambda J^{2}/r^{2}$,
\begin{equation}
  E_{\lambda} \left( r, \mathcal{E}, J \right) = E_{\lambda} \left( \frac{2\lambda |J|}{r}, \mathcal{E}, J \right).
  \label{eqn:spectrum-symmetry}
\end{equation}

There are many approaches to dealing with the second branch of the square root and also the imaginary energies.
One approach is to stick to the negative branch and truncate the spectrum to the real energies.
This involves a loss of locality and also a modification of the defining equation, as emphasised in \cite{Iliesiu:2020zld}.
Another approach is to try to use the second branch to get rid of the imaginary energy states as in \cite{Iliesiu:2020zld}.
In this paper we will, in section \ref{ssec:maxent-ttbar}, find a role for both branches as well as the imaginary energies.

\subsection{Duality with the Bulk}
One of the reasons for the interest in $T \overline{T}$-deformed theories in recent years has been the duality with finite-cutoff holography \cite{McGough:2016lol,Kraus:2018xrn,Kraus:2022mnu}.
Taking the seed CFT to be a holographic CFT dual to pure GR at low energies, the $T \overline{T}$-deformed theory is dual to the bulk with Dirichlet boundary conditions at finite cutoff.
This has been extended in an important way recently \cite{Caputa:2020fbc,Araujo-Regado:2022gvw}, by taking the finite-cutoff surface to be a Cauchy slice.

In this version of the duality, the partition function $Z_{\lambda}[g_{\mu\nu}]$ of $T\overline{T}$ may be reinterpreted as a wavefunction of the metric
\begin{equation}
  \Psi[g_{\mu\nu}] = Z_{\lambda} [g_{\mu\nu}].
\end{equation}
Here, $g$ is a Euclidean two-dimensional metric and we include the counterterm that implements the shift \eqref{eqn:T-shift} in $Z_{\lambda}$.
The ADM momentum operator on this wavefunction is proportional to the stress tensor,
\begin{equation}
  \Pi^{\mu\nu} = -i\frac{\delta}{\delta g_{\mu\nu}} = i\frac{\sqrt{g}}{2} T^{\mu\nu}.
\end{equation}
Then if we rewrite the trace-flow equation \eqref{eqn:tr-flow-eqn} as\footnote{It is important to remember that $\Pi_{\mu\nu} = g_{\mu\alpha} g_{\nu\beta} \Pi^{\alpha\beta} = + i \tfrac{\delta}{\delta g^{\mu\nu}}$.}
\begin{equation}
  0 = \lambda \left[ \left( T^{\mu}_{\mu} \right)^{2} - T^{\mu\nu} T_{\mu\nu} \right] - \frac{c}{24\pi} R - \frac{1}{2\lambda} = \frac{1}{g} \left( \Pi^{\mu\nu} \Pi_{\mu\nu} - \Pi^{2} \right) - \frac{1}{4\lambda} \left( \frac{c}{24\pi} R + \frac{1}{2\lambda} \right)
  \label{eqn:tr-flow-grav}
\end{equation}
and set
\begin{equation}
  \frac{c}{24\pi} = \frac{\ell}{16\pi G_{N}}, \qquad \Lambda = - \frac{1}{\ell^{2}}, \qquad \lambda = 4\pi G_{N} \ell = \frac{\ell^{2}}{c/6\pi}.
  \label{eqn:dictionary}
\end{equation}
we obtain the Hamiltonian constraint for GR with a negative cosmological constant,
\begin{equation}
  H \Psi = \left[-\frac{\ell\sqrt{g}}{16\pi G_{N}}(R - 2\Lambda) + \frac{16\pi  G_{N}}{\ell \sqrt{g}} (\Pi_{\mu\nu}\Pi^{\mu\nu} - \Pi^2) \right] \Psi = 0.
  \label{eqn:ham-cons}
\end{equation}
The same procedure converts the conservation of the stress tensor into the diffeomorphism constraint,
\begin{equation}
  \grad_{\mu} T^{\mu\nu} \propto D^\nu \Psi = \left[-2\grad_\mu \Pi^{\mu\nu}\right] \Psi = 0.
  \label{eqn:Dcon}
\end{equation}

Using Hamilton's equations we can also calculate the extrinsic curvature of a codimension-1 slice $\Sigma$:
\begin{equation}
  K_{\mu\nu} = \frac{1}{2} \mathcal{L}_{n} g_{\mu\nu} = \frac{1}{2} \frac{\partial H}{\partial \Pi^{\mu\nu}} = \frac{16\pi G_{N}}{\sqrt{g}}\left(\Pi_{\mu\nu} - g_{\mu\nu} \Pi \right) = -i \cdot 8\pi G_{N} (T_{\mu\nu} - g_{\mu\nu} T),
  \label{eqn:K-T}
\end{equation}
where the normalisation of the normal covector is always
\begin{equation}
  n_{i} n^{i} = -1.
  \label{eqn:normal-norm}
\end{equation}
This means that $K_{\mu\nu}$ is real ($T_{\mu\nu}$ is imaginary) when the normal is timelike, and imaginary (real) when it is spacelike.

\subsubsection{Adding Bulk Matter} \label{ssec:bulk-matt}
The above equations seem quite reasonable if we start with a CFT which is dual to pure gravity at low energies.
But actually, all of them are true \emph{regardless} of whether the original CFT is holographic; as long as either the metric is flat or $c$ is large.
In fact, the CFT could even be dual to another holographic theory with other light matter fields (e.g. scalars, fermions, gauge fields), and yet paradoxically we would still get the vacuum GR constraint equations \eqref{eqn:ham-cons} and \eqref{eqn:Dcon}, with none of the expected corrections due to matter.\footnote{This was called the `fake bulk' in \cite{Mazenc:2019cfg}.}

However, these equations only concern the equations for the 1-point functions involving $\Pi_{\mu\nu}$ and its powers at a single point.
The theories are not identical at the bulk-quantum level, when we look at higher n-point correlations of $\Pi_{\mu\nu}$ (let alone if we consider other operators).
If we examine the n-point functions, we will presumably discover that they only correspond to an approximately local \emph{bulk} QFT when the undeformed field theory at $\lambda = 0$ is a holographic CFT (with large $c$) dual to pure GR in $D = 3$ (which has no local degrees of freedom).

Hence, if there are other matter fields, additional terms will be needed in \eqref{eqn:ttbar-eqn} in order to obtain the appropriate matter couplings in the constraint equations.
Unfortunately, these additional matter terms will usually result in the deformed theory no longer being exactly solvable except at large $c$, and since the coupling is irrelevant, UV completion may be needed.\footnote{It may seem to introduce a fine-tuning problem that we need to deform the CFT by a specific set of terms, tailored to the particular CFT, in order to obtain the correct bulk theory.
	But this may be looking at things from the wrong perspective.
	If we think of the $T\overline{T}$-deformed theory in the usual Wilsonian manner, it is merely an effective field theory for some UV complete theory which is valid at the Planck scale.
	It may be that this UV theory is naturally local (at least above the Planck length).
	Then, it \emph{happens} to flow to the boundary CFT in the IR.
	From the IR perspective, we shouldn't be surprised if there are constraints on the allowed set of irrelevant deformations needed for locality in the UV, since on a reductionistic perspective the fundamental theory is defined in the UV, not the IR.
	(The only fine-tuning problem that should surprise us from this effective field theory perspective, is the usual fine-tuning issue for the cosmological constant.)
	The other reason not to worry about fine-tuning, is that our results are robust to the details that we fine-tune.
	They depend only on the existence of a local theory in the UV, and on IR quantities matching semi-classical gravity at a perturbative level.
}

\subsection{Holographic Entropy on a Cylinder} \label{ssec:entropy-cyl} In a large-$c$ holographic CFT which is dual to pure gravity at low energies, the microcanonical entropy $S(\mathcal{E},J)$ (i.e. the log of the number of states in a small ${\cal E}$ window with angular momentum $J$) is given at leading order in $c$ by the logarithm of the Cardy entropy \eqref{eqn:cardy-dos}.
It turns out that
\begin{equation}
  \log \rho(\mathcal{E},J) = \frac{A_{+}}{4 G_{N}}.
  \label{eqn:cardy-area}
\end{equation}
where $A_+$ is the area\footnote{Really a length, as we are in D = 3.} of the outer horizon of the dual BTZ black hole.
$J = \mathcal{E}$ corresponds to an extremal BTZ black hole in which the horizons have merged, i.e. $A_+ = A_-$. 
For ${\cal E} < |J| \le {\cal E} + c/24\pi$ the spectrum is sparse (no black hole), while there are no states at all with $|J| > {\cal E} + c/24\pi$.

This Cardy entropy formula \eqref{eqn:cardy-dos} continues to be valid for a $T\overline{T}$-deformed holographic CFT.
This is because it is a form of spectral flow, where eigenstates deform to eigenstates and different energies don't cross along the deformation.
As a result, the density of states at a fixed $\mathcal{E},J$ remains fixed.
Importantly, the density of states at fixed \emph{deformed} energy and angular momentum changes,
\begin{equation}
  \rho_{\lambda} \left( E, J, A \right) = \rho_{0} \left( \mathcal{E} (E,J,A), J \right), \qquad \mathcal{E} (E,J,A) =  \frac{r^{2}}{4\lambda} - \lambda \left( \frac{E}{2\pi} \right)^{2} + \frac{\lambda J^{2}}{r^{2}}.
  \label{eqn:deformed-dos}
\end{equation}
We have defined the area $A = 2\pi r$.
The function $\mathcal{E}(E,J,A)$ is found by inverting \eqref{eqn:deformed-energy}.
Increasing $\lambda/r^{2}$ at fixed $\mathcal{E},J$ is dual to pulling in the boundary in a fixed BTZ geometry as in \cite{McGough:2016lol}, and the invariance of the density of states at fixed $\mathcal{E},J$ is thus the simple geometric fact that the area of the black hole is independent of the position of the boundary, in the microcanonical ensemble.

\begin{figure}[h!]
  \centering
	  \begin{tikzpicture}
    \draw (0,0) -- (0,2) -- (2,0) -- (2,2) -- (0,0);
    \draw (0,2) -- (2,4);
    \draw (2,2) -- (0,4);
    \draw[decorate,decoration={zigzag,amplitude=.3mm,segment length=.5mm}] (2,2) -- (2,4);
    \draw[decorate,decoration={zigzag,amplitude=.3mm,segment length=.5mm}] (0,2) -- (0,4);
    \draw[    blue,thick] (2,0) to[out=135,in=-90] (1.4,1) to[out=90,in=-135] (2,2);
    \draw[red,thick] (2,2) to[out=135,in=-90] (1.4,3) to[out=90,in=-135] (2,4);
    \draw[dotted,thick,->,blue] (1.3,1) -- (1.1,1);
  \end{tikzpicture}
  \caption{Taking $E_{\lambda} \to 0$ is dual to the bounding cylinder approaching the horizon (blue surface). The symmetry \eqref{eqn:spectrum-symmetry} is a relation between cylinders approaching the inner and outer horizons; it takes the blue and red surfaces into each other.}
  \label{fig:old-fin-cut}
\end{figure}
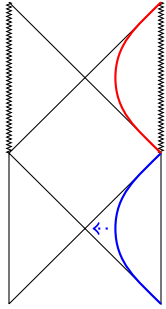

Here we can see the emergence of HRT-like formulas for entropy.
Given an angular momentum $J$ and a radius $r$,  the deformed energy \eqref{eqn:deformed-energy} is real if
\begin{equation}
  0 \le \frac{r^{2}}{\lambda} - 4 \mathcal{E} + 4 \frac{\lambda}{r^{2}} J \quad \implies \quad \mathcal{E} \le \frac{r^{2}}{4\lambda} + \frac{\lambda}{r^{2}} J^{2}.
  \label{eqn:reality-cond}
\end{equation}
Plugging this back into the Cardy formula \eqref{eqn:cardy-dos}, we find that the maximum density of states for a real energy is
\begin{align}
  \log \rho_{0} &\le (2\pi)^{2} \sqrt{\frac{c}{12\pi}} \sqrt{\frac{r^{2}}{4\lambda} + \frac{\lambda J^{2}}{r^{2}} + \sqrt{\left( \frac{r^{2}}{4\lambda} + \frac{\lambda J^{2}}{r^{2}} \right)^{2} - J^{2}} } \nonumber\\
  &\qquad\qquad = (2\pi)^{2} \sqrt{\frac{c}{12\pi}}  \sqrt{\frac{r^{2}}{4\lambda} + \frac{\lambda J^{2}}{r^{2}} + \left|\left( \frac{r^{2}}{4\lambda} - \frac{\lambda J^{2}}{r^{2}} \right)\right| } \nonumber\\
  &\qquad \qquad \xrightarrow{|J| \le \frac{r^{2}}{2\lambda}} (2\pi)^{2} \sqrt{\frac{c r^{2}}{24\pi\lambda}} = \frac{2\pi r}{4 G_{N}}
  = \frac{A}{4 G_{N}}.
  \label{eqn:entropy-bound}
\end{align}
So the entropy is bounded by the area when the deformed energy is real and the angular momentum is subcritical.
In the bulk the limit $E_{\lambda} \to 0$ at fixed $\mathcal{E},J$ is given by the bounding cylinder approaching the horizon as in figure \ref{fig:old-fin-cut}.

The bulk also gives an interpretation of the unexpected symmetry \eqref{eqn:spectrum-symmetry} of the spectrum, also shown in figure \ref{fig:old-fin-cut}.
The subcritical case corresponds to cylinders outside the outer horizon and the supercritical case corresponds to cylinders inside the inner horizon.
In particular, the two horizons map to each other.
It is interesting to consider the above analysis in the supercritical case,
\begin{equation}
  |J| > \frac{r^{2}}{2\lambda} \quad \implies \quad \log \rho_{0} \le \frac{2\pi}{4 G_{N}} \frac{2\lambda |J|}{r}.
  \label{eqn:supcrit-ent-bd}
\end{equation}
If $r$ is the radius of the inner horizon, this entropy continues to be given by the area of the outer horizon.
Inner and outer horizons of a given black hole differ only in the value of $\lambda/r^{2}$; they share the same $\mathcal{E}$ and $J$.
This means that our considerations in this section assign an entropy of $A_{+}/4 G_{N}$ even to the inner horizon.\footnote{Ultimately, the fact that the entropy matches the outer horizon's area, rather than the inner horizon's area, comes from the fact that we are flowing from a normal holographic CFT.
One could attempt to invert this logic, and postulate the existence of a new kind of holographic CFT whose entropy is given by the inner horizon.
In that case, we would need to assign an entropy of $A_{-}/4 G_{N}$ to the outer horizon, by modifying the Cardy entropy formula to
\begin{equation}\label{Sbar}
	\bar{S} \equiv 2\pi \sqrt{\frac{\pi c}{3}
\left({\cal E} - \sqrt{{\cal E}^2 - J^2} \right)} = \frac{\phantom{i}A_-}{4G},
\qquad |J| \le {\cal E},
\end{equation}
where $A_-$ is the area of the inner horizon.
(One can arrive at this by interpreting $E_{\lambda} = 0$ as an equation for $r$ and defining $S = 2\pi r/4 G_{N}$. The equation is quartic so there are four solutions, corresponding to the sign of the two square root radicals in \eqref{eqn:cardy-dos}. The sign of the outer radical is fixed by $S \ge 0$, and the two signs of the inner radical give the lengths of the two horizons.)  While this is an interesting idea, it doesn't seem to correspond to a physically realistic dual CFT.
For it is clear that in a normal CFT, increasing $|J|$ at constant ${\cal E}$ should decrease the number of states.
Hence, $S$ is the correct choice for purposes of counting the states of the black hole, while $\bar{S}$ is not. }
\subsection{The Stress Tensor on a Tube of Non-Constant Radius} \label{ssec:non-const}
In this work, we are interested not just in cylinders but in Cauchy slices, which are never cylinders (due to the asymptotics of AdS).
So, we calculate the one-point function of the stress tensor of the deformed theory on tubes of non-constant radius.
The technique is essentially the same as in \cite{Donnelly:2018bef,Caputa:2019pam,Lewkowycz:2019xse}.
This will be essential in deriving the entropy bound in section \ref{sec:ent-bd}.

We study surfaces whose metric can be written as
\begin{equation}
  ds^{2} = \ell^{2} d\tau^{2} + r(\tau)^{2} d\phi^{2}.
  \label{eqn:tube-g}
\end{equation}
We call these sorts of manifolds ``tubes'' throughout this paper.
In the special case of constant radius, $\partial_{\tau} r = 0$, we call them cylinders.
These tubes have a rotational symmetry, and we will assume that the one-point functions respect this symmetry, which should be satisfied if the boundary conditions at the ends of the tube do not break this symmetry.
See section \ref{app:convs} for further discussion of this assumption in the CFT; the same discussion applies to Cauchy slice boundaries $\partial \Sigma$ which lie at the asymptotic boundary of AdS${}_{3}$.

To find the stress tensor, we use the conservation and trace-flow equations.
For conciseness of exposition, we define
\begin{equation}
  T^{\tau}_{\tau} = \frac{U(\tau)}{2\lambda},\  T^{\phi}_{\phi} = \frac{V(\tau)}{2\lambda},\  T_{\tau\phi} = i \frac{\ell J(\tau)}{r(\tau)}.
  \label{eqn:T-defs}
\end{equation}
The factor of $i$ in $T_{\tau\phi}$ is included to ensure that $J$ is a real angular momentum in real time $t = i \tau$.
The conservation equations simplify to
\begin{equation}
  \grad_{\mu} T^{\mu}_{\tau} = 0 \ \implies \ V = \frac{\dot{\left( r U \right)}}{\dot{r}}, \qquad \grad_{\mu} T^{\mu}_{\phi} = 0 \ \implies \ \dot{J} = 0,
  \label{eqn:cons-eqn}
\end{equation}
where the dot denotes differentiation with respect to $\tau$.\footnote{
	The first equation is the famous pressure relation $T^{\phi}_{\phi} = d \left( r T^{\tau}_{\tau} \right)/dr$.
	While the pressure is definitionally the response of the energy on a cylinder to a change in the radius, here we see that the same relation also holds when the radius changes along a tube.
}
The trace-flow equation simplifies to
\begin{equation}
  U V = 1 + \frac{\lambda c}{6\pi} \frac{R}{2} - \frac{4 \lambda^{2} J^{2}}{r(\tau)^{4}}.
  \label{eqn:tr-flow-eqn-2}
\end{equation}

Plugging in the curvature
\begin{equation}
  \frac{R}{2} = - \frac{1}{\ell^{2}} \frac{\ddot{r}}{r}.
  \label{eqn:tube-R}
\end{equation}
and the conservation equations \eqref{eqn:cons-eqn} into the trace-flow equation \eqref{eqn:tr-flow-eqn} and multiplying both sides by $2 r \dot{r}$ results in
\begin{align}
  \dot{\left( r^{2} U^{2} \right)} &= 2 r \dot{r} - 2 \frac{\lambda c}{6\pi \ell^{2}} \dot{r} \ddot{r} - \frac{8 \lambda^{2} J^{2}}{r^{3}} \dot{r} \nonumber\\
  &= \partial_{\tau} \left( r^{2} - \frac{\lambda c}{6\pi \ell^{2}} \dot{r}^{2} + \frac{4 \lambda^{2} J^{2}}{r^{2}} \right).
  \label{eqn:tr-flow-soln-diff}
\end{align}
This can be integrated to
\begin{equation}
  U^{2} = 1 - \frac{4 \mathsf{C} \lambda + \tfrac{\lambda c}{6\pi \ell^{2}} \dot{r}^{2}}{r^{2}} + \frac{4 \lambda^{2} J^{2}}{r^{4}}.
  \label{eqn:tr-flow-soln}
\end{equation}

There are two ways to fix the integration constant $\mathsf{C}$.
The first is to take the limit $r^{2}/\lambda \to \infty$.
In this limit, we should recover the CFT stress tensor after undoing the shift \eqref{eqn:T-shift}.
We find\footnote{We've picked a sign for the square root so that this limit doesn't diverge.}
\begin{equation}
  \mathcal{T}^{\tau}_{\tau} = \frac{1 - \sqrt{U^{2}}}{2\lambda} \xrightarrow{\lambda \to 0} \frac{\mathsf{C} + \tfrac{c}{24\pi\ell^{2}} \dot{r}^{2} }{r^{2}}.
  \label{eqn:old-T-lim}
\end{equation}
We can compare this to the CFT formula.
The tube is Weyl-related to a cylinder by \eqref{eqn:weyl}, and the stress tensor on this cylinder is given by \eqref{eqn:cyl-T-cft}.
Using the Weyl transformation formula \eqref{eqn:T-weyl} for the stress tensor of a CFT, we find that the stress tensor on the metric \eqref{eqn:tube-g} is
\begin{equation}
  \mathcal{T}^{\tau}_{\tau} = \frac{\mathcal{E} + \tfrac{c}{24\pi\ell^{2}} \dot{r}^{2}}{r^{2}}, \qquad \mathcal{T}^{\phi}_{\phi} = - \frac{\mathcal{E} + \tfrac{c}{24\pi\ell^{2}} \dot{r}^{2}}{r^{2}} + \frac{c}{12\pi\ell^{2}} \frac{\ddot{r}}{r}.
  \label{eqn:cft-T}
\end{equation}
This gives us $\mathsf{C} = \mathcal{E}$, i.e.
\begin{equation}
  U^{2} = 1 - \frac{4\lambda \mathcal{E} + \dot{r}^{2}}{r^{2}} + \frac{4 \lambda^{2} J^{2}}{r^{4}},
  \label{eqn:U-final-tube}
\end{equation}
where we have used the dictionary \eqref{eqn:dictionary} to set $\tfrac{\lambda c}{6\pi} = \ell^{2}$.
The factor multiplying $\dot{r}^{2}$ is actually $\ell^{2} g^{\tau\tau}$, which we have set to $1$ by hand using the freedom to rescale $\tau$.

\begin{figure}[h!]
  \centering
  \begin{subfigure}[c]{.45\textwidth}
		\centering
		    \begin{tikzpicture}
    \draw (0,0) arc (180:315:1 and .5) -- (3,-.35);
    \draw (0,-2) arc (180:45:1 and .5) -- (3,-1.65);
    \draw[dashed] (3, -.35) -- (3.5, -.35);
    \draw[dashed] (3,-1.65) -- (3.5,-1.65);
    \draw (0,0) arc (90:270:.2 and 1);
    \draw (1.71,-.35) arc (95:265:.17 and .65);
    \end{tikzpicture}
    \label{fig:junction-1}
  \end{subfigure}
  \begin{subfigure}[c]{.45\textwidth}
		\centering
		    \begin{tikzpicture}
      \draw (0,0) -- (0,-2) -- (2,0) -- (2,-2) -- (0,0) -- (2,2) (2,0) -- (0,2);
      \draw[decorate,decoration={zigzag,amplitude=.3mm,segment length=.5mm}] (0,0) -- (0,2);
      \draw[decorate,decoration={zigzag,amplitude=.3mm,segment length=.5mm}] (2,0) -- (2,2);
      \draw[darkbrown,thick] (0,-1.2) to[out=0,in=-150] (1.5,-1) to[out=90,in=-135] (2,0);
      \draw[darkbrown,thick] (0, -.2) to[out=0,in=-150] (1.5, 1) to[out=90,in=-135] (2,2);
      \draw[brown,thick] (0, -.8) to[out=0,in=-120] ( 1,-.5) to[out= 0,in=-135] (2,0);
    \end{tikzpicture}
    \label{fig:junction-2}
  \end{subfigure}
  \caption{A junction in the boundary and the bulk. Left: An example tube attached to a constant-radius cylinder. Right: Three ways of embedding the cylinder with a junction into the rotating BTZ black hole spacetime. In all three cases, the constant-radius part of the junction follows a constant-$r$ slice of the bulk. The dark brown slices are spacelike as long as the radius varies and timelike once we set the radius to constant; $\overline{E}$ is real at the junction in this case.
The light brown slice is spacelike everywhere; $\overline{E}$ is imaginary at the junction in this case.}
  \label{fig:junction}
\end{figure}
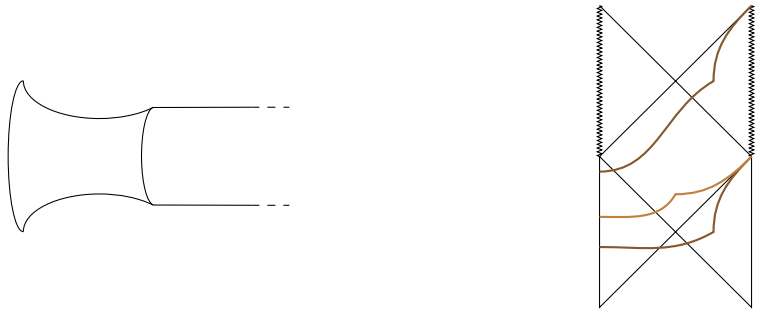

The other way to fix the integration constant $\mathsf{C}$ is to imagine cutting off the tube on one side and pasting in a constant-radius cylinder of radius $r_{0}$ at $\tau = 0$ as in figure \ref{fig:junction}.
At linear order in $\tau$, the radius function is
\begin{equation}
  r(\tau) = r_{0} + \delta v \tau \theta(\tau) + \mathcal{O} \left( \tau^{2} \right).
  \label{eqn:funnel-r}
\end{equation}
For $\tau < 0$, $U^{2}$ takes the form \eqref{eqn:deformed-energy}
\begin{equation}
  r_{0}^{2} U^{2} (0^{-}) = r_{0}^{2} - 4 \lambda \mathcal{E} + \frac{4 \lambda^{2} J^{2}}{r_{0}^{2}}.
  \label{eqn:cyl-T}
\end{equation}
We integrate \eqref{eqn:tr-flow-soln-diff} to obtain the value on the other side of the junction,
\begin{equation}
  r^{2} U^{2} (0^{+}) = r_{0}^{2} - 4 \lambda \mathcal{E} - \delta v^{2} + \frac{4 \lambda^{2} J^{2}}{r_{0}^{2}}.
  \label{eqn:funnel-int}
\end{equation}
This again gives \eqref{eqn:U-final-tube}.

An interesting outcome of this second analysis is that the combination $r^{2} U^{2} + \dot{r}^{2}$ is constant across the junction.
This suggests that we can define an angle $\Theta$ such that
\begin{equation}
  E = \frac{\pi}{\lambda} r U = \overline{E} \cos \Theta, \qquad \frac{\pi}{\lambda} \dot{r} = \overline{E} \sin \Theta.
  \label{eqn:Theta}
\end{equation}
At $\Theta = 0$, i.e. for a constant-radius cylinder, $\overline{E}$ is just the energy.\footnote{
		This representation reveals something potentially useful.
		When we write the trace-flow equation in terms of the function $\Theta(\tau)$, we find that an instantaneous pulse taking $\Theta$ to $\Theta + \pi$ only results in a $\delta$-function in $V$ and leaves $\overline{E}$ unaffected.
		This pulse takes $E \to - E, \dot{r} \to - \dot{r}$ and corresponds to flipping the direction of the normal $n^{(0)}$.
		These sign-flips are exactly what we expect at the junction from the ket copy of the Cauchy slice to the bra copy in a density matrix.
}
More generally,
\begin{equation}
  \overline{E} = \pm \sqrt{E^{2} + \left( \frac{\pi}{\lambda} \dot{r} \right)^{2}}
  \label{eqn:E-bar}
\end{equation}
is the energy on a constant radius cylinder that we transition to using a junction.
Remember that on a Cauchy slice $E \in i \mathds{R}$ and so $\overline{E}$ can be both real and imaginary.
The distinct cases in a rotating BTZ black hole are shown in figure \ref{fig:junction}.
The fact that $\overline{E}$ can be both real and imaginary for Cauchy slices will be the reason that we will advocate for keeping the imaginary energy levels as shown in figure \ref{fig:cross}.

To understand this better, consider a circle at fixed $\tau$, which is codimension-two in the three-dimensional bulk.
This codimension-two surface has two normals and therefore two extrinsic curvatures.
One normal can be taken to be to point along the tube, i.e. $n^{(\tau)} = \ell d\tau$, and the corresponding extrinsic curvature is
\begin{equation}
  K_{(\tau)\phi\phi} = \frac{r \dot{r}}{\ell} \implies K_{(\tau)}  = \frac{\dot{r}}{\ell r}.
  \label{eqn:s1-K-tau}
\end{equation}
The other normal can be taken to be pointing outside the tube, $n^{(0)} = \ell dx^{0}$ with $g_{00} = - \ell^{2}$.
Using \eqref{eqn:K-T}, we have
\begin{equation}
  K_{(0)} = - \frac{2\lambda}{\ell} i \left[ T^{\phi}_{\phi} - \left( T^{\tau}_{\tau} + T^{\phi}_{\phi} \right) \right] = i \frac{2\lambda}{\ell} T^{\tau}_{\tau} = i \frac{U}{\ell}.
  \label{eqn:s1-K-o}
\end{equation}
These two extrinsic curvatures form a covector on the two-dimensional normal space,
\begin{equation}
  K_{\alpha} dx^{\alpha} = \frac{1}{\sqrt{\gamma}} dx^{\alpha} \partial_{\alpha} \sqrt{\gamma} = \ell K_{(0)} dx^{0} + \ell K_{(\tau)} d\tau,
  \label{eqn:K-vec}
\end{equation}
where $\gamma$ is the induced metric on the codimension-two surface.
Its norm
\begin{equation}
  K^{2} \equiv \sum_{\alpha,\beta = 0,\tau} g^{\alpha\beta} K_{\alpha} K_{\beta} =  \left( K_{(\tau)} \right)^{2} - \left( K_{(0)} \right)^{2}  = \frac{\dot{r}^{2} + (rU)^{2}}{\ell^{2} r^{2}}
  \label{eqn:codim-two-ext}
\end{equation}
is independent of coordinate choices on this normal space, depending only on the choice of codimension-two surface.
And it turns out that the quantity $\overline{E}$ is the same as this:
\begin{equation}
  \overline{E}^{2} = \left( \frac{r}{ 4 G_{N} } \right)^{2} K^{2}.
  \label{eqn:bar-E-meaning}
\end{equation}
So, the sign of $\overline{E}^{2}$ tells us about whether the extrinsic curvature vector is spacelike, timelike or null.

To summarise the results of this subsection, the stress tensor on the tube is given by
\begin{align}
  T^{\tau}_{\tau} &= \pm \frac{1}{2\lambda} \sqrt{1 - \frac{4\lambda \mathcal{E} + \dot{r}^{2}}{r^{2}} + 4 \frac{\lambda^{2} J^{2}}{r^{4}}} \nonumber\\
  T^{\phi}_{\phi} &= \pm \frac{1}{2\lambda} \frac{1 - \frac{\lambda  c}{6\pi} \frac{\ddot{r}}{r} - \frac{4 \lambda^{2} J^{2}}{r^{4}}}{\sqrt{1 - \frac{4\lambda \mathcal{E} + \dot{r}^{2}}{r^{2}} + 4 \frac{\lambda^{2} J^{2}}{r^{4}}}} \nonumber\\
  T_{\tau\phi} &= i \frac{\ell J}{r}.
  \label{eqn:tube-T}
\end{align}
Here, we should choose the same sign on the first two lines.

\section{The Holographic Covariant Entropy Bound} \label{sec:bound-gen}
An entropy bound is always the answer to the question: given some data $\mathcal{D}$, what is the maximum entropy of some system $R[\mathcal{D}]$ consistent with the given data?
For this work, we are interested in the case where the given data $\mathcal{D}$ is the data on a codimension-two surface $\sigma$ and the system $R[\mathcal{D}]$ is a subsystem $R$ of the dual CFT.
Apart from the data on the codimension-two surface, the bound also depends on the bulk effective theory.
In this work, we take the theory to be three-dimensional general relativity with negative cosmological constant.

We begin in section \ref{ssec:cft-bd} by illustrating the main logic of this work in a 2d CFT, and apply it to the Cauchy slice theory in section \ref{ssec:cauchy-bd}.
We then provide an alternative upper bound in section \ref{ssec:bulk-bd}, which agrees with the HCEB in the appropriate limit, but is more convenient in many contexts.

\subsection{Warm Up: A Quasi-Local Entropy Bound in 2d CFT} \label{ssec:cft-bd}
Before tackling the main subject of this paper, we illustrate our logic in 2d CFTs, explicitly drawing parallels with some trivial statements about tensor networks.
The 2d CFT story is similar in spirit to \cite{Almheiri:2024,Verlinde:2022xkw,Chandra:2023dgq}, though the perspective is slightly different.

We begin by reminding the reader of a familiar fact about tensor networks, namely the Swingle bound \cite{Swingle:2009bg} on entanglement entropy.
A state in a bipartite Hilbert space $\mathcal{H}_{L} \otimes \mathcal{H}_{R}$ described by a tensor network has the special property that the entanglement between the two tensor factors is bounded by the log of the total bond dimension of the minimal cut that separate the open legs corresponding to $L$ from those of $R$,
\begin{equation}
  S_{E} \le \min_{\mathrm{cuts}\, \gamma} \sum_{l \in \gamma} \log d_{l}.
  \label{eqn:tn-ent}
\end{equation}
It is trivial to write down a quasi-local entropy bound from here: for all cuts $\gamma$ that separate $L$ from $R$, we have
\begin{equation}
  S_{E} \le S_{\mathrm{bd}} (\gamma) = \sum_{\l \in \gamma} \log d_{l}.
  \label{eqn:tn-ent-bd}
\end{equation}

This is often interpreted in terms of a max-flow being bounded by a min-cut \cite{Cui:2015pla,Freedman:2016zud}.
The basic idea is to ``turn the diagram by $90^{\circ}$'' and interpret it as a (non-unitary) quantum circuit.
The flow we seek to maximise is the flow of information through this circuit, and the min-cut is the information bottleneck.
Given the data that the TN contains $\gamma$, either $\gamma$ is this bottleneck or it is not; in the first case, $S_{\mathrm{bd}} (\gamma)$ is the tighest bound on EE we can write down without knowledge of the tensors, and in the second case $S_{\mathrm{bd}} (\gamma)$ upper bounds the min-cut.
Either way, it is an upper bound on the information flow.
The aim of our work is to find a similar quantity for codimension-two surfaces in spacetime.

\begin{figure}[h!]
  \centering
		\begin{tikzpicture}[scale=.6]
		\draw (0,0) rectangle (1,2);
		\draw (3,0) rectangle (4,2);
		\draw node (a) at ( .5,1) {\small $M_{L}$};
		\draw node (b) at (3.5,1) {\small $M_{R}$};

		\foreach \y in {.2,.4,.6,.8,1,1.2,1.4,1.6,1.8}
		{
			\draw (\y - 2.3, 2.5) -- (\y - 2.3,\y) -- (0  ,\y);
			\draw (4,\y) -- (6.3-\y,\y) -- (6.3-\y,2.5);
		}
		\foreach \y in {.5,1,1.5}
		{
			\draw (1,\y) -- (3,\y);
		}

		\draw[dashed] (2,-.5) -- node[below,pos=0] {$\gamma$} (2,2.5);
		\draw[dashed] (-2.4,2.5) -- node[above,pos=0] {$L$} (-.2,2.5);
		\draw[dashed] ( 4.2,2.5) -- node[above,pos=1] {$R$} (6.4,2.5);

		\draw[dotted,->] (0,3) node[above] {\tiny information flow} to[out=-45,in=180] (2,2) to[out=0,in=-135] (4,3);
	\end{tikzpicture}
  \caption{The entanglement in a tensor network state is bounded by the bond dimension of its minimal cut.}
  \label{fig:simple-tn}
\end{figure}
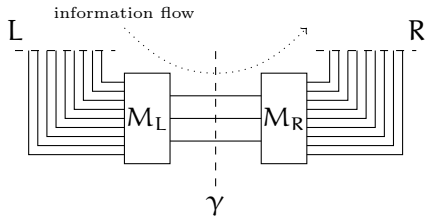

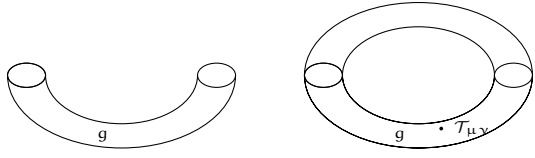
\begin{figure}[h!]
  \centering
		\begin{tikzpicture}[yscale=cos(50)]
		\draw ( 0,0) arc (180:360:1.5);
		\draw (.5,0) arc (180:360:  1);
		\draw (0,0) arc (180:540:.25);
		\draw (0,0) arc (180:540:.25);
		\draw (2.5,0) arc (180:540:.25);
		\node (g) at (1.25,-1.25) {\tiny $g$};
	\end{tikzpicture}
	\qquad
		\begin{tikzpicture}[yscale=cos(50)]
		\draw ( 0,0) arc (180:720:1.5);
		\draw (.5,0) arc (180:720:  1);
		\draw (0,0) arc (180:540:.25);
		\draw (0,0) arc (180:540:.25);
		\draw (2.5,0) arc (180:540:.25);
		\node (g) at (1.25,-1.25) {\tiny $g$};
		\node[circle,fill,inner sep=.5pt,label=right:{\tiny $\mathcal{T}_{\mu\nu}$}] (T) at (1.8,-1.1) {};
	\end{tikzpicture}
  \caption{Left: A CFT state-preparation path integral.\\
	Right: We bound the information flow using the stress tensor one-point function in the norm path integral.}
  \label{fig:cft-pi}
\end{figure}

Before going into the bulk, we consider the example of a holographic 2d CFT.
This is a simpler example where we don't have direct access to an information-theoretic quantity like the bond dimension.
We study a state $\ket{\Psi}_{LR}$ of two 2d CFTs prepared by a Euclidean path integral with arbitrary rotationally-invariant metric and no operator insertions, as shown in figure \ref{fig:cft-pi}.
We can choose coordinates in which the metric takes the form \eqref{eqn:tube-g}.

This preparation path integral does not start out with a useful notion of local bond dimension: the Hilbert space on any constant-$\tau$ cut is infinite-dimensional.
But we can find a bound on the entanglement entropy between the two CFTs with some correlation function data; the tightness of the bound depends on how much correlation function data we have.
Here, we focus on the case where we have the one-point functions of the stress tensor in the closed manifold that calculates the norm $\braket{\Psi}{\Psi}$.

In the neighbourhood of a surface $\tau = \tau_{0}$, the metric is Weyl-equivalent to a cylinder metric
\begin{equation}
  ds^{2} = \frac{r(\tau)^{2}}{\ell^{2}} \left( \ell^{2} d \mathsf{t}^{2} + \ell^{2} d\phi^{2} \right) \equiv e^{2 \Omega(\mathsf{t})} \widehat{ds}^{2}, \qquad d \mathsf{t} \equiv \frac{\ell}{r(\tau)} d\tau.
  \label{eqn:weyl}
\end{equation}
For a holographic CFT without any other sources, we can assume that the one-point function on the manifold with metric $\widehat{ds}^{2}$ must be of the form \eqref{eqn:cyl-T-cft}, as explained near that equation.

To find the maximum entropy that can flow through $\tau = \tau_{0}$, we have to maximise
\begin{align}
  S_{\mathrm{max}} &= \max_{f,\lambda_{1},\lambda_{2},\lambda_{3}} \sum_{J'} \int d\mathcal{E}' \rho(\mathcal{E}',J') f(\mathcal{E}',J') \log \frac{\rho(\mathcal{E}',J')}{f(\mathcal{E}',J')}  + \lambda_{1} \left[ \sum_{J'} \int d\mathcal{E}' \rho(\mathcal{E}',J') f(\mathcal{E}',J') - 1 \right] \nonumber\\
  &\qquad + \lambda_{2} \left[ \sum_{J'} \int d\mathcal{E}' \rho(\mathcal{E}'J') \mathcal{E}' f(\mathcal{E}'J') - \mathcal{E} \right] + \lambda_{3} \left[ \sum_{J'} \int d\mathcal{E}' \rho (\mathcal{E}',J') J' f(\mathcal{E}',J') - J \right].
  \label{eqn:ent-maxing}
\end{align}
Here, the three Lagrange multipliers impose normalisation of the state and the correct expectation values for $\mathcal{E},J$.
Extremising with respect to $f$, we find the equation
\begin{equation}
  \rho \left( \log \frac{\rho}{f} - 1 + \lambda_{1} + \lambda_{2} \mathcal{E}' + \lambda_{3} J' \right) = 0 \quad\implies \quad f = \rho\, \exp( \lambda_{1} + \lambda_{2} \mathcal{E}' + \lambda_{3} J' - 1 )
  \label{eqn:maxent-saddle}
\end{equation}
Imposing the constraints then gives us the grand canonical state, i.e. the thermofield double with temperature $\beta_{*}$ and rotation chemical potential $\mu_{*}$ such that the one-point functions above are reproduced, 
\begin{align}
	\ket{\Psi, \mathrm{MaxEnt}} = M_{L} M_{R} \frac{1}{\sqrt{Z(\beta_{*},\mu_{*})}} \sum_{J} \int \dd{\mathcal{E}} \sqrt{\rho(\mathcal{E},J)}\, &\exp({- \frac{\beta_{*}}{2\pi\ell} \mathcal{E} + \frac{\beta_{*} \mu_{*}}{2\pi \ell} J}) 
  \label{eqn:max-ent-cft} \\
															&\qquad \quad \ket{\mathcal{E},J}_{\tau_{0} - \delta\tau} \ket{\mathcal{E},J}_{\tau_{0} + \delta\tau}.
															\nonumber
\end{align}
Here, $\rho(\mathcal{E},J)$ is the Cardy density of states \eqref{eqn:cardy-dos} and $M_{L}, M_{R}$ are non-unitary operators on the two sides that implement the rest of the path integral.
While this analysis only shows that this is a \emph{local} maximum, we can show also that it is a \emph{global} maximum using the concavity of von Neumann entropy.\footnote{
	Let us briefly outline the proof.
	Consider the reduced density matrix $\rho$ of one of the CFTs.
  Suppose that $\rho$ is a local maximum of the von Neumann entropy and $\rho'$ is a global maximum with larger entropy and the same expectation values.
  Then any density matrix $t \rho + (1-t) \rho'$ also has the same expectation values.
  By concavity, $S [t \rho + (1-t) \rho'] \ge t S(\rho) + (1-t) S(\rho') \ge S(\rho)$, where the last inequality is saturated only at $t=0$.
  As we send $t \to 0$, the entropy approaches $S(\rho)$ from above, contradicting the assumption that $\rho$ is a local maximum of the entropy.
}

The entanglement entropy between $L$ and $R$ in this state is, at leading order in $c$, simply
\begin{equation}
  S_{E,\mathrm{max}} = \log \rho \!\left( \mathcal{E}, J \right).
  \label{eqn:max-ent}
\end{equation}
Since the microcanonical version of \eqref{eqn:max-ent-cft} has the same entropy at this order, we could as well consider the microcanonical state
\begin{equation}
  \ket{\Psi, \mathrm{Micro}} = M_{L} M_{R} \frac{1}{\sqrt{\rho(\mathcal{E},J)}} \int_{\mathcal{E} - \delta \mathcal{E}}^{\mathcal{E} + \delta \mathcal{E}} d\mathcal{E}' \ket{\mathcal{E}',J}_{\tau_{0} - \delta\tau} \ket{\mathcal{E}',J}_{\tau_{0} + \delta \tau}, \qquad \delta \mathcal{E} = \mathcal{O} \left( c^{0} \right)
  \label{eqn:max-ent-mucan}
\end{equation}
and find the same entropy bound at leading order.
The microcanonical state would be the maximum entropy state if we were also given that the variance in the energy is $\mathcal{O} (c^{0})$.  Given only the one-point functions, it is only a good approximation.  (When we generalise this argument to the $T \overline{T}$-deformed theory in section \ref{sssec:ttb-ensembles}, we will find that the canonical ensemble is no longer defined.  Then we \emph{have} to fix the fluctuations to be small, since the maximum entropy given only one-point functions diverges.)

By a similar logic as in the tensor network case,
\begin{equation}
  S_{E} \le \min_{\tau} S_{E,\mathrm{max}} (\tau) \le S_{E,\mathrm{max}} (\tau).
  \label{eqn:cft-bd}
\end{equation}
Thus, \eqref{eqn:max-ent} is a quasi-local entropy bound in 2d CFT with our assumptions.
\cite{Chandra:2023dgq} also deals with the rotationally non-symmetric case.

This same MaxEnt procedure works when pasting the original path integral preparing $\ket{\Psi}$ to one preparing $\ket{\Phi}$.
The one-point functions of the stress tensor in the path integral calculating $\braket{\Phi}{\Psi}$ may take different values from those in $\braket{\Psi}{\Psi}$ in general.
The arguments above indicate that \eqref{eqn:max-ent} is a bound on the pseudo-entropy \cite{Nakata:2020luh} between $L$ and $R$.

\subsection{An Entropy Bound in Cauchy Slice Holography} \label{ssec:cauchy-bd}
We now use a similar MaxEnt procedure not for 2d CFTs themselves but for a CFT state pasted to a $T \overline{T}$-deformed 2d CFT.
We then interpret the circle $\tau = \tau_{0}$ as a codimension-two slice of a bulk spacetime and interpret the MaxEnt as an entropy flow through the bulk surface.
The conceptual leap here is that we interpret \emph{every} codimension-two surface in the bulk spacetime as admitting an infinite-dimensional Hilbert space like that of a CFT, instead of an $\exp(A/4 G_{N})$-dimensional one.
The bound is equal to area sometime, but even then it is not the (log-)dimension of the Hilbert space, but a quasi-local entropy bound of the sort exhibited above.

Consider a state $\ket{\Psi}$ in the boundary CFT.
By the Cauchy slice holography dictionary, its WdW wavefunction is given by an inner product between the state $\ket{\Psi}$ and a state prepared by the Cauchy slice path integral, as shown in \eqref{eqn:csh}.
For concreteness, we take the Cauchy slice theory to be the original $T \overline{T}$-deformed theory with only stress-tensor deformations --- in other words, we take the bulk theory to be pure GR.
Most of our considerations will generalise straightforwardly to other dimensions and the case with bulk matter, as we will argue in \cite{Soni:2024} in more detail.
We will also assume rotational symmetry here for definiteness, but again the main arguments generalise away from this situation, and the explicit bound in pure GR will not require this restriction.

Consider a $T \overline{T}$-deformed partition function on a rotationally-symmetric 2d manifold with metric $g$, ending at circles of radius $r_{1}$ and $r_{2}$.
This partition function calculates
\begin{equation}
  Z_{\lambda} [g, F] = \sum_{J_{1}, J_{2}} \int d\mathcal{E}_{1} d\mathcal{E}_{2}\, F(\mathcal{E}_{1}, J_{1}, \mathcal{E}_{2}, J_{2}) \mel{\mathcal{E}_{1},J_{1},r_{1}, \dot{r}_{1}}{\mathds{T} [g]}{\mathcal{E}_{2}, J_{2}, r_{2}, \dot{r}_{2}},
  \label{eqn:transition-matrix}
\end{equation}
where the function $F$ encodes the boundary conditions.
Although the placement of $r_{1,2}, \dot{r}_{1,2}$ indicate that they are labels defining the states, they are actually labels defining the Hilbert space $\mathcal{H}_{r,\dot{r}}$.
The spectrum \eqref{eqn:tube-T} exhibits that the Hilbert space actually depends not only on $r_{1,2}$ but also on $\dot{r}_{1,2}$.

The discussion around figure \ref{fig:junction} defines a natural unitary
\begin{align}
  \tilde{\mathds{J}}: \mathcal{H}_{r,\dot{r}} &\to \mathcal{H}_{r,0} \nonumber\\
  \tilde{\mathds{J}} \ket{E,J} &= \ket{\overline{E} = E \sec \Theta , J},
  \label{eqn:junct-map}
\end{align}
where $\Theta$ was defined in \eqref{eqn:Theta}.
We can use the junction to define such a map in other theories as well \cite{Araujo-Regado:2022gvw}.
There is also a natural unitary between $\mathcal{H}_{r_{1},0} \to \mathcal{H}_{r_{2},0}$, which maps states of the same $\mathcal{E},J$ to each other.

In the limit $r \to \infty$, the Hilbert space $\mathcal{H}_{r,0}$ just becomes the CFT Hilbert space on an infinitely large circle.
This can be related to the CFT Hilbert space on a circle of radius $r_{0}$ using a dilatation operator.
Denote by $\mathds{J} (r_{0},r)$ the composition of $\tilde{\mathds{J}}$ and the scaling.
The $T \overline{T}$-deformed partition function, in the limit $r_{1,2} \to \infty$, can as well be written as\footnote{In the main text we are assuming that $\Sigma$ is smooth.
But if it has sharp corners, one may also need to insert the junction conditions at intermediate points here.}
\begin{align}
	Z_{\lambda} [g,F] = \sum_{J_{1,2}} \int d^{2} \mathcal{E}\, & F(\mathcal{E}_{1,2}, J_{1,2}) \nonumber\\
	 &\lim_{r_{1,2} \to \infty} \mel{\mathcal{E}_{1},J_{1},r_{1}^{(0)}}{\mathds{J} (r_{1}^{(0)}, r_{1})  \mathds{T} [g] \mathds{J} (r_{2}^{0}, r_{2})^{\dagger}}{\mathcal{E}_{2}, J_{2}, r_{2}^{(0)}}.
  \label{eqn:transition-cft}
\end{align}

The function $F$ is nothing but the wavefunction of the CFT state, and so we may interpret the partition function above as
\begin{equation}
  Z_{\lambda} [g, \Psi] = \braket{g}{\Psi},
  \label{eqn:csh-dict}
\end{equation}
where
\begin{equation}
  \bra{g} \hat{P} \equiv \sum_{J_{1,2}} \int d^{2} \mathcal{E}\, \mel{\mathcal{E}_{1}, J_{1}}{\mathds{J}_{L} \mathds{T} [g] \mathds{J}_{R}^{\dagger}}{\mathcal{E}_{2}, J_{2}} \bra{\mathcal{E}_{1}, J_{1}} \bra{\mathcal{E}_{2}, J_{2}}
  \label{eqn:wdw-state}
\end{equation}
is an analog of the fixed-metric Wheeler-DeWitt ``state,'' projected onto the physical CFT Hilbert space ($\hat{P}$ is the projector).
The Cauchy slice holography dictionary states that $\Psi[g]$ is the Wheeler-DeWitt wavefunction of the CFT state $\ket{\Psi}$.
This is an explicit realisation of the bulk-boundary map defined in \cite{Araujo-Regado:2022gvw}.

Assuming that $\ket{\Psi}$ is prepared by some CFT path integral, $\braket{g}{\Psi}$ can be interpreted as a path integral obtained by gluing the CFT state-preparation path integral to that of the deformed theory on $g$.
The value of the stress tensor at $\sigma$ read off from the bulk extrinsic curvatures is the expectation value in this glued path integral.
Let us assume that $\ket{\Psi}$ has $\mathcal{O} (G_{N}^{0})$ fluctuations in simple expectation values for simplicity; there is no loss of generality here, since we can project onto this case without changing the classical geometry.
With this projection, the argument of section \ref{ssec:cft-bd} tells us that $S_{\mathrm{bd}} (\sigma)$ is a bound on the log of the rank of the pseudo-density matrix $\tr_{L} \left[ \ket{\Psi} \bra{g} \hat{P} \right]$.

\begin{figure}[h!]
  \centering
	\includegraphics{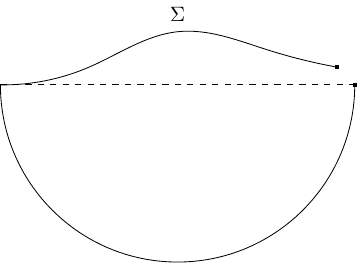}
  \caption{The HCEB is a bound on the rank of the pseudo-density matrix defined by the path integral between the two dots.}
  \label{fig:pseudo-rho}
\end{figure}

\subsubsection{From Pseudo-Entropy to Entanglement Entropy}
The rank of the pseudo-density matrix is not in general an upper bound on the rank of the true density matrix $\tr_{L} \ket{\Psi} \bra{\Psi}$, which  is the object that actually controls the entanglement entropy between the $L,R$ systems in the state $\ket{\Psi}$.
We now argue that, for semi-classical states, a bound on the rank of the pseudo-density matrix $\tr_{L} \left[ \hat{P} \ket{g} \ket{\Psi} \right]$ is also (approximately) a bound on the rank of the true density matrix.
We argue that this is the case if the Cauchy slice with metric $g$ can be embedded in the Lorentzian bulk dual of $\ket{\Psi}$.
We provide a second argument in appendix \ref{app:eq-2}.
We define $\ket{\Psi}$ to be normalised, $\braket{\Psi}{\Psi} = 1$.

The argument proceeds as follows.
Given the extrinsic curvature on a Cauchy slice in a spacetime dual to $\ket{\Psi}$, we can take a superposition of \eqref{eqn:wdw-state} to make a coherent state $\hat{P} \ket{g,T}$.
Here, we have used the dictionary \eqref{eqn:K-T} to translate the stress tensor to an extrinsic curvature.
This is a semi-classical state with small fluctuations in the metric and extrinsic curvatures, such that the product of the uncertainties is $\mathcal{O} \left( G_{N}^{0} \right)$.\footnote{
	For a general discussion on coherent states, see \cite{Yaffe:1981vf}.
	In particular, we only demand classicality, which means that the uncertainties in conjugate pairs need not be equal.
}
We can create such a state for any Cauchy slice.
See \cite{Hartnoll:2022snh} for a recent explicit construction in the minisuperspace approximation.

The introduction of coherent states serves two purposes. The first is to ensure that we are reading off extrinsic curvatures of the spacetime dual to $\ket{\Psi}$. In general, $\braket{g}{\Psi}$ need not be dominated by the same spacetime as $\braket{\Psi}{\Psi}$, even if the latter contains a slice with metric $g$.
The other purpose it serves is that there could be multiple Cauchy slices with metric $g$ in the same spacetime.\footnote{
	An example is if the bulk dual has time-reversal symmetry: the two Cauchy slices have the same extrinsic curvatures with different signs and the expectation value of the stress tensor would be $0$ even if the right geometry dominates in the WdW wavefunction $\braket{g}{\Psi}$.
}
These will generically have different extrinsic curvatures and so the corresponding coherent states will be different.

Since $\hat{P} \ket{g,T}$ lives in the CFT Hilbert space, we can ask which state it is.
It is a state that has the same bulk geometry as $\ket{\Psi}$.
For example, if $\ket{\Psi}$ is the canonical TFD, $\hat{P} \ket{g,T}$ could be a microcanonical TFD.
By the HRT formula,\footnote{
	We don't need to use the HRT formula here; all we need is that the entanglement entropy is determined at leading order by the classical geometry.
	This can be possibly established by using a de Finetti theorem to argue that the modular Hamiltonian of the CFT state is a classical operator \cite{Magan:2017udh}.
	This would imply the following: since $\ket{\Psi}$ and $\hat{P} \ket{g,T}$ differ only at the quantum level, the entanglement entropy of both states must be the same at leading order.
	This comment is relevant if we would like to use the HCEB to independently re-derive the HRT formula.
}
\begin{equation}
  S_{R} (g,T) = S_{E} (\Psi)
  \label{eqn:entropy-equal}
\end{equation}
at leading order.

\subsection{A Quasi-Local Entropy Bound in Gravity} \label{ssec:bulk-bd}
Let us now define an entropy bound in semi-classical gravity.
This is equal to the above in the semi-classical limit, but this will sometimes be a more convenient statement.

Suppose we are given the intrinsic geometry and extrinsic curvature of a $(d-1)$-dimensional surface $\sigma$, as well as the values and derivatives of bulk matter fields on it, along with a guarantee that it can be embedded in a $(d+1)$-dimensional spacetime that satisfies Einstein's equations with cosmological constant $\Lambda = - d(d-1)/2\ell^{2}$ and some given matter content.
In each such permissible spacetime $\mathcal{M}$, there is a boundary region $R(\sigma,\mathcal{M})$ homologous to $\sigma$; there are often two choices for $R$, but there could be just one as in the end-of-the-world brane example shown in figure \ref{fig:ads-vaidya}.\footnote{
	Or even none, if $\sigma$ is in a baby universe.
	In that case $R(\sigma,\mathcal{M})$ is the empty set.
}
Now, there is a minimal extremal surface $X(\sigma,\mathcal{M})$ homologous to $R(\sigma,\mathcal{M})$.\footnote{
	Informally, one can say simply that $X$ is homologous to $\sigma$.
	We avoid saying that because we would like to allow cases where $X,\sigma$ are not spacelike-separated.
	
	For the case of positive cosmological constant, we should extend the definition of homology to allow this.
	In this case, if we don't specify the system $R[\sigma,\mathcal{M}]$ but simply define its entropy to be given by the area of the minimal extremal surface, we can make statements parallel to many below.
}
By the HRT formula,
\begin{equation}
  S_{R(\sigma,\mathcal{M})} = \frac{\mathrm{Area} [X(\sigma,\mathcal{M})]}{4 G_{N}}.
  \label{eqn:S-A}
\end{equation}
A priori, the bound for the entropy flowing through $\sigma$ given the above data is
\begin{equation}
  S_{\mathrm{bd}} (\sigma) = \max_{\mathcal{M} \supset \sigma} S_{R(\sigma,\mathcal{M})}.
  \label{eqn:a-priori-bd}
\end{equation}
The argument in section \ref{ssec:cauchy-bd} implies that, when the boundary state is dominated by a single geometry, this bound agrees with the one found from Cauchy slice holography.

\section{The Radical Nature of the Spectrum}\label{sec:ttb-spect}
The first step in performing the MaxEnt procedure in the deformed theory dual to pure gravity is understanding the spectrum and Hilbert space.
This is what we turn to in this section.
While we study this is the $T \overline{T}$-deformed CFT that is dual to pure GR in the bulk, the main ideas discussed below will generalise also to other bulk theories (including both matter as well as higher-derivative corrections).

\subsection{The Cross-Shaped Spectrum} \label{ssec:maxent-ttbar}

The existence of the branch cut in the spectrum  \eqref{eqn:deformed-energy} raises two urgent interpretational questions:
\begin{enumerate}
\item For all $\mathcal{E} > 0$, the energy eigenvalues move from the real axis to the imaginary axis as we increase $\tfrac{\lambda}{r^{2}}$, signalling a breakdown of unitarity.
Is this part of the spectrum physical, or should it be removed?
\item Relatedly, what is the significance of the fact that the square root in \eqref{eqn:deformed-energy} has \emph{two} possible signs?
\end{enumerate}
We suggest answers to both of these questions below.

\begin{figure}[h!]
  \centering
	  \begin{tikzpicture}[baseline]
    \draw (-1,0) -- (1,0);
    \draw[->] (0,0) -- (0, 3);
    \draw[->] (0,0) -- (0,-3);
    \draw[dotted] (-3,0) -- (-1,0);
    \draw[dotted] ( 3,0) -- ( 1,0);
    \draw (2.5,2.5) -- (2.5,2) -- (3,2);
    \node (E) at (2.8,2.3) {$E$};
    \draw[dotted,thick,brown,->] (-2,.3) -- node[pos=.5,above] {$\mathcal{E}$} (-1,.3);
    \draw[dotted,thick,brown,->] ( 2,.3) -- ( 1,.3);
    \draw[dotted,thick,brown,->] (.3,  .5) -- (.3,  1.5);
    \draw[dotted,thick,brown,->] (.3,-.5) -- (.3,-1.5);
  \end{tikzpicture}
	\qquad\qquad
	  \begin{tikzpicture}[domain=-2:2,baseline=1.5cm]
		\draw[<->] (-2.2,0) -- node[pos=.9,below] {$\abs{E}$} (2.2,0);
		\draw[->] (0,0) -- node[pos=1,above] {$\log \rho(E)$} (0,3);
		\draw[samples=75]      plot (\x,{ sqrt(4 - \x*\x) });
		\node (E) at (2.5,.3) {$E \in \mathds{R}$};
		\draw plot (\x,{ sqrt(4 + \x*\x) }) node[above right] {$E \in i \mathds{R}$};
	\end{tikzpicture}
  \caption{The cross-shaped spectrum (left) and the density of states (right). We have set $J = 0$.}
  \label{fig:def-dos}
\end{figure}
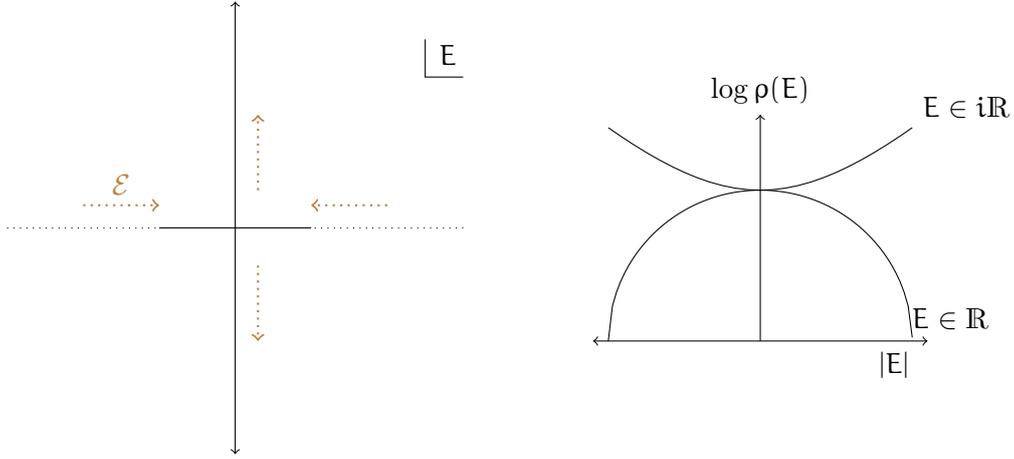

Some authors have advocated for the point of view that the imaginary-$E$ part of the $T\overline{T}$ spectrum is unphysical and should be removed \cite{Iliesiu:2020zld}.\footnote{
	A truncation can also be performed to make the theory finite-dimensional and therefore to make the deformation with matter well-defined \cite{Batra:2024kjl}; this is logically different from truncating all imaginary levels, and is consistent with the perspective we eventually advocate.
}
This would result in a slightly non-local theory.
Since (after fixing $J$ and $r$) there are only a finite number of energy eigenvalues on the real axis, the Hilbert space of a surface area $A$ is finite-dimensional, and the entropy flow in this theory is bounded above by a finite value similar to a discrete tensor network.
If we keep only the real-energy subcritical states, the maximum microcanonical entropy is at $\overline{E} = 0$ as shown in figure \ref{fig:def-dos}, giving a na\"ive entropy bound
\begin{equation}
  S \le S_{\mathrm{neb}} (\sigma) = \frac{A(\sigma)}{4 G_{N}},
  \label{eqn:neb}
\end{equation}
as derived in \eqref{eqn:entropy-bound}.
Saturation happens when the surface $\sigma$ is a marginal or extremal surface.
If we include the supercritical states as well, we find a `duplex' na\"ive entropy bound\footnote{
	The appearance of this dual length suggests that when $|J|$ becomes large enough, perhaps the degrees of freedom are no longer usefully organized in the $\phi$ direction, but rather some other degrees of freedom begin to open up which are not localized in the $\phi$ direction. This has a little bit of the flavor of $T$-duality in string theory where small circles in target space are equivalent to large circles, due to winding modes becoming low energy. With respect to the original $\phi$-direction, the entropy can exceed the Planck density by an arbitrary amount, posing problems for discrete quantum gravity models in which the number of discrete components in the ultraviolet scales with the area alone, like random tensor networks.
}
\begin{equation}
  S \le S_{\mathrm{dneb}} = \frac{\max(A, \overline{A})}{4 G_{N}}, \qquad \overline{A} = 2\lambda \abs{\int d\phi T^{\tau}_{\phi}}.
  \label{eqn:dneb}
\end{equation}
This was derived in \eqref{eqn:supcrit-ent-bd}.

However, neither of these proposals is good enough for our purposes.
For example, by enclosing an object with a slice that wiggles in time, one can make the enclosing area arbitrarily small.
One can also find violations of these na\"{i}ve entropy bounds inside black holes,\footnote{For example, inside a non-rotating BTZ black hole of radius $r_{h}$, $A < 2\pi r_{h}$ and $\overline{A} = 0$.} and in cosmology.\footnote{
	This was one of Bousso's motivations to invent his covariant entropy bound \cite{Bousso:1999xy}, which applies only to lightsheets (i.e. nowhere expanding null surfaces).
}
On a related note, we saw in figure \ref{fig:junction}, surfaces inside a black hole have $\overline{E} \in i \mathds{R}$ and so these proposals cannot apply to them.
So we need to include the imaginary energy states as well.

The full, cross-shaped spectrum is shown in figure \ref{fig:cross} and also in figure \ref{fig:def-dos} below.
As one gradually increases $\lambda/r^{2}$, it is possible for a state to transition across the branch point from the real part of the spectrum to the imaginary part.
Then we have a seemingly arbitrary choice---should we turn right or left at the branch cut?  

If we only allow one imaginary sign ($i$ or $-i$) for $E$, the theory breaks $CPT$ symmetry, since $CPT$ sends $E \to E^*$, $J \to J$.
This seems bad, so we had better allow states with \emph{both} imaginary signs.
But if we allow both signs, then as we adjust $r$ a single state with $E < 0$ discontinuously maps to \emph{two} imaginary states, one with $+i$ sign and the other with $-i$ sign.
And a discontinuous change in the count of states as one slowly changes $r$ \emph{also} seems bad, as it would lead to sharp jumps in the partition function as the source parameters are adjusted.

The only obvious way to fix this, is to also include the extra states on the real axis with $E > 0$.
Then the branch-point transitions involve 2 real eigenvalues $\to$ 2 imaginary eigenvalues.
It follows that we need to include twice as many states in the spectrum when $\lambda > 0$.
To make this explicit, we can write
\begin{equation}\label{cross}
  E = \pm\frac{\pi}{\lambda} \sqrt{r^{2} - 4\lambda \mathcal{E} + \frac{4 \lambda^2 J^2}{r^2}},
\end{equation}
with the $\pm$ reminding us that there are twice as many states in the $T\overline{T}$ spectrum as in the CFT, since there are two possible choices of sign for the square root.
This gives a cross-shaped spectrum, since there are states on both the real and imaginary axes of the $E$-plane, as in figure \ref{fig:cross}.
There is an $E \to -E$ symmetry of the spectrum, as well as an $E \to E^*$ symmetry (related to CPT).

This doubling can be understood easily using the duality with the bulk.
For a given codimension-one surface in spacetime, there are two possible normals, and the corresponding extrinsic curvatures (and therefore stress tensors) differ by a sign.
This is why the spectrum needs to have an $E \to -E$ symmetry.
In the $\lambda/r^{2} \to 0$ limit, half of these states get gapped out by an infinite amount after we reverse the shift of the energy in \eqref{eqn:T-shift} \cite{Freidel:2008sh}.

In appendix \ref{app:price}, we discuss the most natural inner product on this Hilbert space, which we find to be (assuming a discrete spectrum), proportional to
	\begin{equation}
		\braket{E_{1}}{E_{2}} = \delta_{E_{2},E_{1}^{*}}
		\begin{cases}
			-\operatorname{sgn} E_{2} & E_{2} \in \mathds{R} \\
			1 & E_{2} \in i \mathds{R}
		\end{cases}
		\label{eqn:ip-summary}
	\end{equation}
The new states with the $+$ sign in \eqref{cross} approach, as $\lambda/r^{2} \to 0$, a CFT-like spectrum where the energy is bounded \emph{above}.

The real and imaginary parts of the spectrum are quite different from each other, since $S$ monotonically increases as a function of ${\cal E}$.
On the real $E$-axis, after fixing $J$ and $r$, there are only finitely many real states, with a minimum and maximum energy given by:
\begin{equation}
E_\text{max} = - E_\text{min} = 
\sqrt{\frac{r^2 + \lambda c/6\pi}{\lambda^2} + \frac{4J^2}{r^2}},
\end{equation}
The density of states as a function of $E$ at $J=0$, using the algorithm explained in \eqref{eqn:deformed-dos}, is
\begin{equation}
  \log \rho_{\lambda} (E,J=0,A) = 2\pi \ell \sqrt{\frac{\pi^{2} r^{2}}{\lambda^{2}} - E^{2}}.
  \label{eqn:deformed-dos-explicit}
\end{equation}
This is shown in figure \ref{fig:def-dos}.
The maximum microcanonical entropy for real energies is at $E = 0$ with $S = A_+/4 G_{N}$, as discussed in section \ref{ssec:entropy-cyl}.
However, on the imaginary $E$-axis there are infinitely many states, and they have a Hagedorn-like spectrum (in the sense that the number of states grows exponentially with $|\Im(E)|$ as $\Im(E) \to \pm \infty$).

In the bulk dual, the limit $r^{2}/\lambda \to 0$ corresponds to approaching the \emph{singularity} of a black hole (of increasingly large horizon radius, if we keep $r$ fixed).
Since the model is exactly solvable for flat cylinders, it seems one can get arbitrarily close to the singularity; this may not be very significant however, since this singularity is not a curvature singularity.

\subsection{Bulk Interpretation of the Spectrum} \label{ssec:x-shape}

These extra imaginary-energy states we are including in the spectrum are, as already mentioned, crucial for constructing a holographic theory of Cauchy slices in a Lorentzian-signature spacetime.
It is worth dwelling on this point further.

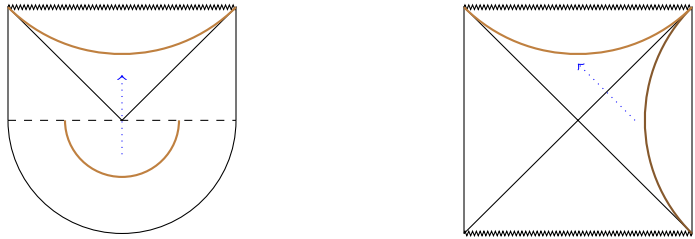
\begin{figure}[h!]
  \centering
	  \begin{tikzpicture}[scale=1.5]
    \draw (0,1) -- (0,2) -- (1,1) -- (2,2) -- (2,1);
    \draw[dashed,thin] (0,1) -- (2,1);
    \draw[decorate,decoration={zigzag,amplitude=.3mm,segment length=.5mm}] (0,2) -- (2,2);
    \draw (0,1) arc (180:360:1);
    \draw[brown,thick] (.5,1) arc (180:360:.5);
    \draw[brown,thick] (0,2) to[out=-45,in=-135] (2,2);
    \draw[dotted,blue,->] (1,.7) -- node[pos=.7,right] {} (1,1.4);

    \begin{scope}[shift={(4,0)}]
    \draw (0,0) -- (0,2) -- (2,0) -- (2,2) -- (0,0);
    \draw[decorate,decoration={zigzag,amplitude=.3mm,segment length=.5mm}] (0,2) -- (2,2);
    \draw[decorate,decoration={zigzag,amplitude=.3mm,segment length=.5mm}] (0,0) -- (2,0);
    \draw[darkbrown,thick] (2,0) to[out=135,in=-135] (2,2);
    \draw[brown,thick] (0,2) to[out=-45,in=-135] (2,2);
    \draw[dotted,blue,->] (1.5,1) -- node[pos=.7,right] {} (1,1.5);
    \end{scope}
  \end{tikzpicture}
  \caption{Two ways of transitioning from real to imaginary energy in Schwarzchild-BTZ, by decreasing the radius of the cylinder to a value less than than the horizon radius, $r < r_{h}$. Left: keep the metric Euclidean but change the signature of the ambient spacetime. Right: Change the signature of the metric and cross the horizon.
}
  \label{fig:im-trans}
\end{figure}

In the case of BTZ, the states on the imaginary axis become important on a Cauchy slice that passes into the interior (black hole or white hole) regions of the BTZ spacetime, as shown in figure \ref{fig:junction}.\footnote{
In general this phenomenon will not be confined to the interior of event horizons, as there are also surfaces with real $K_{\mu\nu}$ outside of horizons, even in vacuum Minkowski or AdS.
But such surfaces are either noncompact, or else are not flat cylinders.
Hence it is easiest to explore these issues on a BTZ spacetime.}
There are two possible ways to think about the transition from real to imaginary energies depending on whether we also change the signature of the cylinder metric (see figure \ref{fig:im-trans}):  
\begin{itemize}
\item If we choose to keep the metric of the cylinder Euclidean on both sides of the transition, then the real-$E$ states correspond to slices of Euclidean-BTZ, while the imaginary-$E$ states correspond to Lorentzian-BTZ inside the horizon.

\item If on the other hand, we wish to always remain on the Lorentzian-BTZ geometry, then we need to take the cylinder to have Lorentzian signature outside the horizon (real-$E$), but Euclidean signature inside the horizon (imaginary-$E$).
	This is similar to the analysis of \cite{Caceres:2022smh,Hartnoll:2022snh,Blacker:2023oan,Blacker:2023ezy,Blacker:2024rje}.
\end{itemize}
In fact, this feature of the spectrum resolves a perplexity about black hole thermodynamics which one of the authors (ACW) has suffered from ever since he was a graduate student, which is how to give a statistical interpretation of trapped surfaces ${\cal T}$ inside of a black hole.
Inside of the black hole there are 1-parameter families of null separated trapped surfaces with decreasing $A_{\cal T}$.
These seem to violate the Second Law of Thermodynamics if we assume $S_{\cal T} = A_{\cal T}/4 G_{N}$.
One could avoid thinking about this perplexity by asserting that there is only a state-counting interpretation on the event horizon, but we want to understand the microscopic statistics of quantum gravity everywhere, not just at a horizon!

But if we quantize a trapped surface in BTZ by thinking of it as a slice of a stationary cylinder inside of the horizon, with $A_{\cal T} < A_+$, and we look at the $T\overline{T}$ theory on that cylinder, we find that these trapped surfaces inside the black hole have imaginary $E$ and hence actually $S_{\cal T} > A_{\cal T}/4$.
So we should think of them as having more entropy than their area, resolving the tension with thermodynamics.

\subsection{Thermal States}\label{sssec:ttb-ensembles}

This unusual cross-shaped spectrum creates a problem with the MaxEnt procedure that we need to define our entropy bound.
For example, consider a density matrix of the deformed theory on a circle that is an equal mix of two microcanonical ensembles at $\overline{E} = \pm i \alpha$, for some $\alpha > 0$.
The entropy of this state diverges as we send $\alpha \to \infty$.
However, the expectation value of $\overline{E}$ in this state is $0$ no matter the value of $\alpha$!

This is because the usual definition of the (grand) canonical ensemble is highly problematic, as there are infinitely many states on the imaginary axis, which would contribute an oscillatory sum to $Z = \tr(e^{-\beta H})$, even at fixed $J$.
If we vary over $J$ values, things are even worse, as $E$ would no longer be bounded below.
This does not signal vacuum instability since $J$ is conserved, but it is not behavior one would expect in a normal field theory!

Yet there is no problem defining the microcanonical inverse temperature $\beta_{\mathrm{eff}} = (\partial S/\partial E)|_J$ on the support of the spectrum.
This of course means that expectation values in the canonical ensemble can be calculated by saddle-point (but the precise meaning of such calculations needs to be considered carefully).
We find
\begin{equation}
  \beta_{\mathrm{eff}} = - 2\lambda E \frac{\sqrt{\mathcal{E} (E,J)^{2} - J^{2}}}{2\pi \sqrt{\pi c/3}}.
  \label{eqn:mucan-temp}
\end{equation}
This microcanonical $\beta_{\mathrm{eff}}$ can take any real value including $\pm \infty$ (corresponding to the states with minimum or maximum $\Re E$), or it can take imaginary values.
The two branches are joined at $E = 0$ where $\beta = 0$; this was used to derive a version of \eqref{eqn:neb} in \cite{Caputa:2020fbc}.

Thus, if we have $E = 0$, the microcanonical temperature is $\beta_{\mathrm{eff}} = 0$, and the canonical ensemble is a uniform mixture over the entire Hilbert space, leading to a divergent entropy.
But $\overline{E} = 0$ exactly when $\sigma$ is marginal or extremal, which are the two cases in which $S \le A/4 G_{N}$ has actually been proved \cite{Lewkowycz:2013nqa,Dong:2016hjy,Dong:2017xht,Engelhardt:2018kcs}!
Even when $\beta_{\mathrm{eff}} > 0$, the canonical partition function diverges because the Boltzmann weight becomes $\exp( i \beta_{\mathrm{eff}} |E| )$ for large $\mathcal{E}$.
Finally, if $E \in i \mathds{R}$, we find that $\beta_{\mathrm{eff}}$ and $E$ have opposite phases, making $\beta_{\mathrm{eff}} E > 0$; this leads to a Boltzmann suppression of large $E$, but in this regime the density of states as a function of $E$ has a Hagedorn growth and so the partition function still diverges.

How can this be?
The answer lies in the fluctuations of $\overline{E}$.
The states where $\ev{\overline{E}} = 0$ but $S > A/4 G_{N}$ are ones where the variance $\ev{\overline{E} \overline{E}} - \ev{\overline{E}}^{2}$ is large; we have not proved this statement but it is clear in our example.
Whereas the proofs mentioned above assume a classical spacetime, which means that the extrinsic curvatures of any surface has fluctuations that vanish as $G_{N} \to 0$.

Hence, to define a sensible ensemble, we must perform a slightly different entropy maximisation than the one in section \ref{ssec:cft-bd}: we not only fix the one-point functions of the stress tensor but also demand that it has small fluctuations.
We call this small fluctuations criterion classicality, since it is dual to the existence of a single bulk geometry.
The MaxEnt procedure with this data results not in the canonical but in the microcanonical ensemble.
This gives
\begin{equation}
	S_{\mathrm{bd}} (\sigma) = \log \rho_{\lambda} \left[ \overline{E} (\sigma), A (\sigma), J(\sigma) \right] = \log \rho_{0} \left[ \mathcal{E} \left( \overline{E}, J, A \right) \right].
  \label{eqn:S-sigma-abstract}
\end{equation}
Notice that, even though we also mentioned the microcanonical ensemble in section \ref{ssec:cft-bd}, there it was an approximation we \emph{could} make at leading order, whereas here we are \emph{forced} to demand classicality.

Actually, the microcanonical ensemble is only the MaxEnt state if fluctuations in the energy scale as $o (G_{N}^{0})$.
We can allow for larger fluctuations, for example $\mathcal{O} ( G_{N}^{-1/2} )$ --- which corresponds to $\mathcal{O} ( G_{N}^{1/2} )$ fluctuations in the extrinsic curvature.
In that case, we can use the fact that there is another possible definition of a thermal state, where the thermal factor is $\exp( - \beta \mathcal{E} / 2\pi r )$.
This is the state we get by deforming a thermal state in the CFT, and it does not suffer the pathologies pointed out above.
In the limit we are working in, the difference is subleading.

\subsection{Mock Unitarity of Euclidean Evolution}
Another interesting feature of a spectrum with imaginary energies is that, when acting on states with imaginary energy, the Euclidean evolution operator acts as
\begin{equation}
  e^{- H \tau} \ket{i E} = e^{- i E \tau} \ket{i E}.
  \label{eqn:phase}
\end{equation}
This looks like the action of a unitary operator in that it only acts as a phase, but it is not actually unitary, i.e. $U^{\dagger} U \neq 1$, under the inner product of appendix \ref{app:price}.
However, some properties that we normally associate to unitarity are exhibited by Euclidean evolution in this case; most importantly, the probability amplitude of individual energy eigenstates are unchanged under Euclidean evolution.

Further discussion of this spectrum, specifically related to the sense in which the Hamiltonian can still be understood as self-adjoint, can be found in appendix \ref{app:spectrum}.

\section{The Entropy Bound for Pure GR} \label{sec:ent-bd}
We now derive the explicit bound for pure GR in the bulk.  In section \ref{ssec:entropy}, we work out the explicit bound for an arbitrary codimension-two surface $\sigma$ with the topology of a circle embedded in a solution of 3d GR with negative cosmological constant.
We then look at specific types of surfaces in section \ref{ssec:egs} and discuss its relation to other entropy bounds in section \ref{ssec:other-bds}.

\subsection{The Entropy of an Arbitrary Surface} \label{ssec:entropy}
Consider a codimension-two surface $\sigma$ in a three-dimensional asymptotically AdS${}_{3}$ spacetime that solves Einstein's equations without matter.
We require that $\sigma$ has the topology of a circle, and that it is homologous to an asymptotic boundary of AdS.
We assume that we know the intrinsic metric and the extrinsic curvatures.
The entropy we assign to it is as follows.

\paragraph{Rotationally Symmetric Surfaces}
We first deal with the rotationally symmetric case.
Consider a Cauchy slice $\Sigma \supset \sigma$ that is rotationally symmetric.
Concretely, if we define $\sigma$ as $\tau = 0$, the metric on $\Sigma$ must take the form \eqref{eqn:tube-g}, and the stress tensor must take the form \eqref{eqn:tube-T}.\footnote{
	While the results of section \ref{ssec:non-const} were worked out for the case when the full manifold has the metric \eqref{eqn:tube-g}, it is easy to check that the results apply even when the metric has that form for an open interval in $\tau$.
	In particular, they apply even if there is classical matter in the spacetime as long as the stress tensor vanishes on this open interval; however, as we will see in \cite{Soni:2024}, this does not mean the entropy bound is un-modified.
}

The extrinsic curvatures that can be associated to a one-dimensional surface in three bulk dimensions are the expansions and the twist.
Consider two future-pointing null normals $k,l$ of $\sigma$ such that $k\cdot k = l \cdot l = 0, k \cdot l = -1$.
We use the letters $k,l$ to denote both the vectors and the dual co-vector.
We also demand that the inaffinities of $k$ and $l$ vanish, i.e.\footnote{If we extend $k,l$ to vector fields, then it is not in general possible to impose both of these conditions everywhere; but it is possible to choose $k,l$ so that this is true at a \emph{particular} codimension-two surface.
}
\begin{equation}
  k \cdot \grad k^{i} = l \cdot \grad l^{i} = 0.
  \label{eqn:inaffinity}
\end{equation}
Then the expansions and the twist are defined as
\begin{equation}
  \theta_{(k)} = \grad^{\phi} k_{\phi}, \qquad \theta_{(l)} = \grad^{\phi} l_{\phi}, \qquad \chi_{\phi} = \frac{1}{2} l^{i} \grad_{\phi} k_{i} = - \frac{1}{2} k^{i} \grad_{\phi} l_{i}.
  \label{eqn:exps}
\end{equation}

To relate these to the results of section \ref{ssec:non-const}, we define the timelike and spacelike normal co-vectors
\begin{equation}
  n_{(0)}  = \frac{k + l}{2}, \qquad n_{(\tau)} = \frac{k - l}{2}.
  \label{eqn:non-null-normal}
\end{equation}
The same equations hold for the corresponding normal vectors.
It is easy to check $-n_{(0)} \cdot n_{(0)} = n_{(\tau)} \cdot n_{(\tau)} = 1, n_{(0)} \cdot n_{(\tau)} = 0$.
We take $\Sigma$ such that, on $\sigma$, $n_{(0)}$ is normal to $\Sigma$ and $n_{(\tau)}$ tangent to it.
To connect to the results of section \ref{ssec:non-const}, we take $d\tau = \tfrac{1}{\ell} n_{(\tau)}$.
With these choices, the components of the extrinsic curvature on $\Sigma$ are
\begin{align}
  K_{(0)} \equiv K\indices{_{(0)}^{\phi}_{\phi}} &= \grad^{\phi} n_{(0)\,\phi} = \frac{\theta_{(k)} + \theta_{(l)}}{2} \nonumber\\
  K_{(0)\tau\phi} &= \grad_{\phi} n_{(0)\,\tau} = \ell n^{i}_{(\tau)} \grad_{\phi} n_{(0)\,i} = \frac{\ell}{4} \left( k^{i} \grad_{\phi} l_{i} - l^{i} \grad_{\phi} k_{i} \right) = -\frac{\ell}{2} \chi_{\phi}.
  \label{eqn:K-theta-reln}
\end{align}
In the penultimate equality, we have used $k^{i} \grad_{\phi} k_{i} = \tfrac{1}2 \grad_{\phi} (k \cdot k) = 0$ and similarly with $l$.
It is also useful to note that
\begin{align}
  K_{(\tau)} = \grad^{\phi} n_{(\tau)\,\phi} = \frac{\theta_{(k)} - \theta_{(l)}}{2}.
  \label{eqn:exp-rp}
\end{align}
So the quantity $K^{2}$ defined in \eqref{eqn:codim-two-ext} is just
\begin{equation}
  K^{2} =  \left( K_{(\tau)} \right)^{2} - \left( K_{(0)} \right)^{2}  = - 2 \theta_{(k)} \theta_{(l)}.
  \label{eqn:K-sq}
\end{equation}

Shifting perspective, when $\Sigma$ is thought of as carrying a $T \overline{T}$-deformed partition function, the stress tensor is related to the same extrinsic curvatures as
\begin{align}
  \overline{E}^{2} &= \left( \frac{r}{4 G_{N}} \right)^{2} K^{2} \nonumber\\
  T_{\tau\phi} = \frac{\ell}{r} i J &= \frac{i}{8\pi G_{N}}  K_{\tau\phi} = - i \frac{\ell}{ 16 \pi G_{N}} \chi_{\phi}.
  \label{eqn:T-K-ent}
\end{align}

Since $\overline{E}$ is the effective energy on a flat cylinder of radius $r$, we can use \eqref{eqn:deformed-energy} to solve for $\mathcal{E}$.
We find
\begin{align}
  J &= - \frac{r}{16 \pi G_{N}} \chi_{\phi} \nonumber\\
  \mathcal{E} &= \frac{1}{4\lambda} \left[ r^{2} - \left( \frac{\lambda \overline{E}}{\pi} \right)^{2} + \left( \frac{2\lambda J}{r} \right)^{2} \right] = \frac{r^{2} \left( 1 - \ell^{2} K^{2} \right) + \ell^{2}\chi_{\phi}^{2}/4}{16 \pi G_{N} \ell}.
  \label{eqn:3d-level-soln}
\end{align}
The entropy we assign to $\sigma$ is the Cardy entropy for the above values $\mathcal{E},J$
\begin{equation}
  S_{\mathrm{bd}} (\sigma) = 2\pi \sqrt{\frac{\pi \ell}{2 G_{N}} \left( \mathcal{E} + \sqrt{\mathcal{E}^{2} - J^{2}} \right)}.
  \label{eqn:S-sigma}
\end{equation}
At a subcritical marginal or extremal surface, manipulations similar to \eqref{eqn:entropy-bound} show that this reduces to $A/4 G_{N}$.

In the case when $J = 0$, the formula simplifies to
\begin{equation}
  S_{\mathrm{bd}} (\sigma) = 2\pi \sqrt{\frac{\pi \ell}{G_{N}} \mathcal{E}} = \frac{1}{4 G_{N}} \sqrt{A^{2} (1 - \ell^{2} K^{2})}.
  \label{eqn:S-sigma-simple}
\end{equation}
This version makes it clear that it reduces to the area when either $\theta_{\pm}$ vanishes.\footnote{
	$J = 0$ means that the dual area vanishes, and so the entropy is the area.
}

\paragraph{Surfaces without Rotational Symmetry}
Let us turn to surfaces $\sigma$ without rotational symmetry.
The crucial facts about the function $S_{\mathrm{bd}}$ that we use are
\begin{enumerate}
  \item It depends only on the homotopy class of $\sigma$. Every homotopically equivalent surface in a solution of pure three-dimensional GR has the same $S_{\mathrm{bd}}$.
  \item It equals the area at an extremal surface.
\end{enumerate}

It is easy to find quantities satisfying the first requirement.
We use the Chern-Simons formulation of three-dimensional gravity \cite{Witten:1988hc,Donnay:2016iyk}.
The metric configuration can be converted to a flat $SO(2,2)$ gauge connection $A_{i}^{\mathtt{a}}$, as detailed in the above references.
Flatness means that $F_{ij}^{\mathtt{a}} = 0$, and that the Wilson loops
\begin{equation}
  W_{R} [\sigma] = \tr_{R} P e^{i \oint_{\sigma} A^{\mathtt{a}} T^{\mathtt{a}}}.
  \label{eqn:wilson-loop}
\end{equation}
are homotopy invariants.
Thus, it is reasonable to expect that the entropy bound is a function of the set of Wilson loops in different irreps $R$,
\begin{equation}
	S_{\mathrm{bd}} (\sigma) = S_{\mathrm{bd}} \left[ \left\{ W_{R} (\sigma) \right\} \right].
  \label{eqn:S-bd}
\end{equation}

We will not write out this function explicitly, but explicit expressions realise this expectation.
\cite{Cangemi:1992my} showed that in a BTZ spacetime the holonomy can be written as
\begin{equation}
  P e^{i \oint_{\sigma} A^{\mathsf{a}} T^{\mathsf{a}}} = V \exp{i \frac{2\pi}{\ell} \left( r_{-} J_{2} + r_{+} P_{2} \right)} V^{-1}, \qquad J_{2}, P_{2} \in \mathfrak{so} (2,2),
  \label{eqn:btz-hol}
\end{equation}
where $V$ depends on the beginning and ending of the integration contour, and cancels out in the Wilson loops.
$r_{\pm}$ are the radii of the outer and inner horizons homotopic to $\sigma$.
The group element between $V,V^{-1}$ is a gauge-invariant quantity.
$S_{\mathrm{bd}}$ is the coefficient of $P_{2}$ times $\ell/4G_{N}$.
This formula has appeared at varying levels of explicitness in \cite{McGough:2013gka,Ammon:2013hba,Castro:2018srf,Mertens:2022ujr,Wong:2022eiu,Akers:2024wab}, though with slightly different interpretations.

Note that this function matches the area of an extremal surface only if there exists an extremal surface $X$ that $\sigma$ can be deformed to.
A more careful proof that this is the right quantity might sometimes require us to construct an auxiliary spacetime containing $\sigma$ where such an extremal surface exists, as in \cite{Engelhardt:2018kcs,Nomura:2018aus,Bousso:2018fou,Bousso:2019dxk}.

\subsubsection{Constancy of $S(\sigma)$ on BTZ} \label{sssec:btz}
Let us show explicitly that this entropy is the same for all rotationally symmetric Cauchy slices of a non-rotating BTZ black hole.
This constancy property is also true for rotating BTZ black holes and for other Cauchy slices, as indicated in \eqref{eqn:btz-hol}, but we will choose a simple case for illustration.

Consider a two-sided non-rotating BTZ black hole in ``Penrose diagram'' coordinates,
\begin{equation}
  ds^{2} = \frac{\ell^{2} \left( - ds^{2} + dy^{2} \right) + r_{h}^{2} \cos^{2} s\, d\phi^{2}}{\cos^{2} y}, \qquad s, y \in \left( - \frac{\pi}{2}, \frac{\pi}{2} \right).
  \label{eqn:btz}
\end{equation}
We define $\sigma$ as the surface $\left\{ s = s_{0}, y = y_{0} \right\}$, and the Cauchy slice $\Sigma$ to be the surface $\left\{ s = s_{0} \right\}$.
The metric on $\Sigma$ is
\begin{equation}
  ds_{\Sigma}^{2} = \frac{\ell^{2} dy^{2}}{\cos^{2} y} + \frac{r_{h}^{2} \cos^{2} s_{0}}{\cos^{2} y} d\phi^{2},
  \label{eqn:btz-cauchy}
\end{equation}
which is of the form \eqref{eqn:tube-g} with $\tau = \int_{0}^{y} \sec y' \, dy' = \sinh^{-1} \tan y$ and $r(\tau) = r_{h} \cos s_{0} \sec y(\tau) = r_{h} \cos s_{0} \cosh \tau$.
The extrinsic curvature tensor and energy are
\begin{equation}
  K_{\mu\nu} = \frac{\cos y}{2\ell} \partial_{s_{0}} g_{\mu\nu} (\Sigma)  = \left(
   \begin{matrix}
       0 & 0\\
       0  & \frac{r_{h}^{2}}{\ell} \frac{\sin s_{0} \cos s_{0}}{\cos y}\\
 \end{matrix}
 \right)
 \quad \implies \quad
 E = \int d\phi\, r(\tau) T^{\tau}_{\tau} = -i \frac{r_{h} \sin s_{0}}{4 G_{N} \ell}.
  \label{eqn:btz-T}
\end{equation}
The other components of the stress tensor vanish.
We also need
\begin{equation}
  \dot{r} = \frac{dy}{d\tau} \frac{d}{dy} \left( \frac{r_{h} \cos s_{0}}{\cos y} \right) = r_{h} \cos s_{0} \tan y.
  \label{eqn:rp-BTZ}
\end{equation}

Plugging these values into \eqref{eqn:3d-level-soln}, we get
\begin{equation}
  4 G_{N} \ell \mathcal{E} = \frac{r_{h}^{2} \cos^{2} s_{0}}{\cos^{2} y} + r_{h}^{2} \sin^{2} s_{0} - r_{h}^{2} \cos^{2} s_{0} \tan^{2} y = r_{h}^{2}.
  \label{eqn:BTZ-level}
\end{equation}
This is a constant, and in fact the correct relation between the CFT dimension and the radius of the black hole.
Plugging this into the Cardy entropy gives $2\pi r_{h}/4 G_{N}$, which is in fact the entropy of the boundary CFT.

An interesting special case of this is $r_{h}^{2} = - \ell^{2}$, which is global AdS.
In this case, the specifics of the analysis above are not valid; for example  the Cauchy slice is not a tube but a disk with metric $\ell^{2} d\tau^{2} + \ell^{2} \sinh^{2} \tau d\phi^{2}$.
However, the result \eqref{eqn:BTZ-level} turns out to be true anyway.
The CFT only has one state at this level, giving an entropy of $0$ ---  as expected for the ground state.

\begin{figure}[h!]
	\centering
		\begin{tikzpicture}
		\draw (4,0) -- (4,4);
		\draw (1.26,1.26) -- (4,4);
		\draw (1.26,2.74) -- (4,0);
		\draw[decorate,decoration={zigzag,amplitude=.3mm,segment length=.5mm}] (2,0) -- (4,0);
		\draw[decorate,decoration={zigzag,amplitude=.3mm,segment length=.5mm}] (2,4) -- (4,4);
		\draw[red,ultra thick] (2,0) to[out=135,in=-135] (2,4);
	\end{tikzpicture}
	\qquad\qquad
	\includegraphics{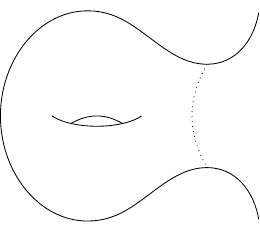}
	\qquad\qquad
		\begin{tikzpicture}
		\draw (0,0) -- (0,4);
		\draw (4,0) -- (4,4);
		\draw (0,0) -- (4,4);
		\draw (0,4) -- (4,0);
		\draw[decorate,decoration={zigzag,amplitude=.3mm,segment length=.5mm}] (0,0) -- (4,0);
		\draw[decorate,decoration={zigzag,amplitude=.3mm,segment length=.5mm}] (0,4) -- (4,4);
		\draw[darkgreen,ultra thick,decorate,decoration={snake,amplitude=.2mm,segment length=.7mm}] (4,.3) -- (.3,4);
	\end{tikzpicture}
	\caption{Examples where the entropy bound is not saturated. Left: A black hole pure state with an end-of-the-world brane behind the horizon. The entropy of the right CFT is $0$ in this case, but the entropy bound is still the area of the horizon. Middle: A pure-state black hole with a toroidal `bag-of-gold' behind the horizon. Right: An AdS-Vaidya geometry. The entropy we associate to a surface to the future of the shock is the area of the larger horizon, which is larger than the entanglement entropy.}
	\label{fig:ads-vaidya}
\end{figure}
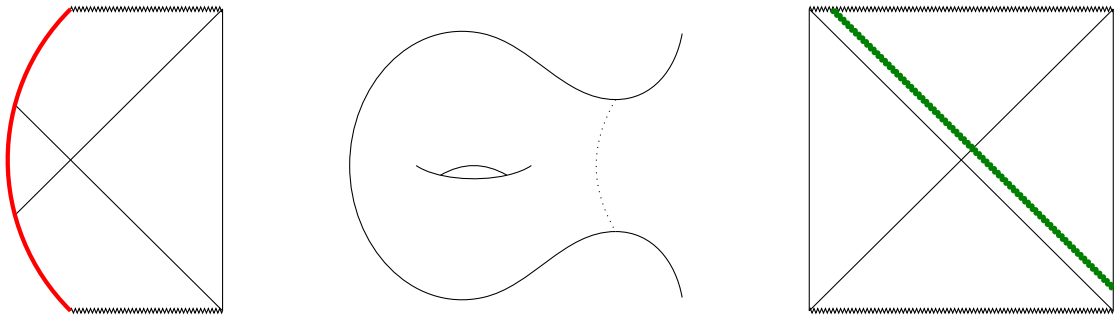

It is important to note that this bound is not always saturated.
A simple example is a locally BTZ solution capped off by an end-of-the-world brane behind the horizon, as shown in figure \ref{fig:ads-vaidya}.
In that case there is only one boundary CFT and it is in a pure state; however, the bound gives the area of the horizon regardless.
Instead of an end-of-the-world brane, the interior could instead contain a higher-genus `bag of gold,' like those discussed in \cite{Krasnov:2000zq,Maloney:2015ina} and references therein; in this case, the bound is the entropy of the horizon for surface homotopic to it but there are also other surfaces wrapping the non-trivial spatial cycles which would have different entropy bounds.
Another example is an AdS-Vaidya type solution, as in figure \ref{fig:ads-vaidya}.\footnote{
	In this case, there is matter in the bulk. We take the considerations of this section to apply in each matter-free region.
	We postpone an in-depth discussion of the theory with bulk matter to \cite{Soni:2024}.
}
In this case, the bound differs across the shock; to the past, we get the actual entanglement entropy between the two CFTs, but to the future we get a larger value, which is the Cardy entropy at the ADM mass of the right CFT after we send in the shock.

\subsection{Various Choices for $\sigma$} \label{ssec:egs}
We now discuss various choices for the surface $\sigma$, namely marginal, extremal and non-marginal surfaces.

\subsubsection{Marginally Trapped Surfaces}
A marginally trapped surface $\mu$ has $K_{i}$ null, since either $\theta_{(k)}$ or $\theta_{(l)}$ vanishes, so $\overline{E} = 0$.
Thus,\footnote{Although we have written $S_\mathrm{CFT}$, this will also hold for the boundary entropy if the boundary is moved in by a finite amount, by another $T\overline{T}$ deformation.}
\begin{equation}
  S_{\mathrm{CFT}} \le S_{\mathrm{bd}} (\mu) = \frac{\max \left( A, \overline{A} \right)}{4 G_{N}},
  \label{eqn:dceb}
\end{equation}
where $\overline{A}$ is the dual area defined as in \eqref{eqn:dneb}.
We need to include the maximisation because the surface $\mu$ could be supercritical.
Hence, the maximum amount of entropy flowing through a subcritical marginally trapped surfaces is equal to its Bekenstein-Hawking formula.
An interesting consequence is that the second law for holographic screens \cite{Bousso:2015qqa} is also a second law for the HCEB.

One might wonder whether it is really necessary to account for the supercritical case.
It is common not to worry too much about pathologies on the inner horizon, due to the expectation that in a model which includes matter, the inner horizon is generically unstable due to blue-shifted infalling modes.
At the pure GR level, there is no matter to be blue-shifted, and so this resolution is not available to us.
Furthermore, as the entropy bound varies continuously on a classical spacetime, it is possible to see the effect by considering surfaces that are close to the inner horizon but don't cross it --- such surfaces would remain even if the inner horizon is replaced by a null singularity.
In \cite{Soni:2024} we will show that the analysis is modified by the presence of matter, but in an even more extreme way: there is no finite entropy bound inside the outer horizon.

\subsubsection{Extremal Surfaces}
As a special case of a marginally trapped surface, one can consider the case of an extremal surface $X$, for which the vector $K_i = 0$.
From the above analysis it follows that
\begin{equation}
  S_\text{CFT} \le \frac{\max(A, \overline{A})}{4G_{N}},
\end{equation}
which looks sort of like the famous HRT formula for holographic entropy.
However, the HRT formula is an equality, not an inequality; and it involves $A$ alone, not $\overline{A}$.

Regarding the issue of $\overline{A}$, the HRT formula is usually applied only in globally hyperbolic regions of black hole spacetimes that exclude the inner horizons of Kerr black holes.
Otherwise HRT is not actually valid.
The reason this has to be an inequality is simply that this entropy bound uses data localised to a single codimension-two surface.
The RHS in the HRT formula is the area of the \emph{globally minimal} extremal surface, whereas the HCEB results in the area of the extremal surface it is evaluated on, which might be a non-minimal extremal surface.
This is the case, for example, in the one-sided black holes of figure \ref{fig:ads-vaidya}.

\subsubsection{Nonmarginal surfaces}

Of course, one of the main novelties of this work is that our results extend beyond marginal and extremal surfaces.
For an untrapped (sometimes called ``normal'') surface $\Upsilon$ with spacelike $K_{i}$, $\overline{E} \in \mathds{R}$ and the maximum allowed entropy at a given $J, r$ is less than the area,
\begin{equation}\label{untrap}
  S_\text{CFT} \le S_{\mathrm{bd}} (\Upsilon) < \frac{\max(A, \overline{A})(\Upsilon)}{4G_{N}}.
\end{equation}
Conversely, for a trapped surface $\cal T$ with timelike $K_{i}$, $\overline{E} \in i \mathds{R}$ and the maximum allowed entropy \emph{exceeds} the area,
\begin{equation}\label{trap}
  S_{\mathrm{bd}} ({\cal T}) > \frac{\max(A, \overline{A})(\mathcal{T})}{4G_{N}}.
\end{equation}
However, unless we specify the rest of the Cauchy slice $\Sigma$, we cannot necessarily say the same about $S_\mathrm{CFT}$, since $S_{\mathrm{CFT}} \le S_{\mathrm{bd}} (\mathcal{T})$ and so we cannot combine these inequalities.

Let us now calculate $S_{\mathrm{bd}}(\sigma)$ more precisely.
In terms of the dual radius
\begin{equation}
  \overline{r} = \frac{2\lambda J}{r}, \qquad \overline{A} = 2\pi \overline{r},
  \label{eqn:rbar}
\end{equation}
the bound is
\begin{equation}
  S_{\mathrm{bd}} = \frac{2\pi}{4G_{N}} \sqrt{\frac{\left[ r^{2} \left( 1 - \ell^{2} K^{2} \right) + \overline{r}^{2} \right] + \sqrt{\left[ r^{2} \left( 1 - \ell^{2} K^{2} \right) - \overline{r}^{2} \right]^{2} - 4 \ell^{2} r^{2} \overline{r}^{2} K^{2}}}{2}}.
  \label{eqn:S-sig}
\end{equation}

To learn something from this rather unedifying expression, we expand this in $K^{2}$ to look at the behaviour close to a marginal or extremal surface.
We find\footnote{
	In the extremal case, where $A = \overline{A}$ on the horizon, we find $S_{\mathrm{bd}} \approx A \left[ 1 - K^{2}/(-2\Lambda) \right]/4 G_{N}.$}
\begin{equation}
  S_{\mathrm{bd}} = \frac{\max(A, \overline{A})}{4 G_{N}} \left[  1 - \frac{A^{2}}{\abs*{A^{2} - \overline{A}^{2}}} \frac{K^{2}}{(-2\Lambda)} \right].
  \label{eqn:S-bd-pert}
\end{equation}
Note the dependence on the square of the extrinsic curvature $K^2$, which is positive in the spacelike case and negative in the timelike case, agreeing with the inequalities \eqref{untrap} and \eqref{trap}.
This expression also extends to the case $\Lambda > 0$ (we will derive it carefully in section \ref{ssec:hawking-mass}).
In this case, the inequalities \eqref{trap} and \eqref{untrap} are reversed, connecting to the results of \cite{Shyam:2021ciy,Coleman:2021nor,Batra:2024kjl}.

\subsection{Relation to Other Entropy Bounds} \label{ssec:other-bds}
\subsubsection{The Bousso Bound} \label{sssec:bousso}

Any given surface $\sigma$ has two future-null expansions $\theta_{(k)}$ and $\theta_{(l)}$.
When one of these expansions is nonpositive, say $\theta_{(l)} \le 0$, then if you extend out a null surface $N$ from $\sigma$ in the direction of the $l$ normal, the null energy condition implies that $\theta_{(l)} \le 0$ on the entire surface.
(This assumes we truncate $N$ when lightrays intersect.)
Such a surface $N$ is called a lightsheet.
According to Bousso's original covariant entropy bound \cite{Bousso:1999xy}, the entropy of fields falling across $N$ is bounded by
\begin{equation}
S_N \le \frac{A(\sigma)}{4G_N}.
\end{equation}

Unfortunately this bound was not totally well defined in QFT due to the nonlocality of quantum field excitations.
Various extensions of this bound which take quantum effects into account are discussed in \cite{Bousso:2014sda} and \cite{Bousso:2015mna}.
These proposed modifications are somewhat orthogonal to the direction of this paper.
In this paper we are ignoring such quantum effects.

Our new HCEB looks qualitatively similar to Bousso's covariant bound in the case of a marginal surface $\mu$.
Both bounds have the implication that an entropy associated with a marginally trapped surface $\mu$ must be less than its area.
However, there are three key differences:
\begin{enumerate}
	\item The HCEB is not a bound on the entropy of bulk matter falling across a codimension 1 null surface $N$, but rather a bound on the UV entropy flowing through the codimension 2 surface $\sigma$ itself, and hence the entropy $S_{\mathrm{CFT}}$ of any homologous region.
		The null surface $N$ is irrelevant.
	\item The bound is equal to the area only when $\theta_{(l)} = 0$, not when $\theta_{(l)} < 0$.
		In the case where $\theta_{(k)} \theta_{(l)} < 0$, the Bekenstein-Hawking formula still provides an upper bound albeit a weaker one.
		But the Bekenstein-Hawking formula is too \emph{strong} in the trapped case $\theta_{(k)} \theta_{(l)} > 0$ (this is not in conflict with the Bousso bound since the light rays end \cite{Bousso:2022cun}).
	\item In the supercritical case, the dual area $\overline{A}$ replaces the normal area $A$.
\end{enumerate}
It is important to emphasize that these modifications are not mix-and-match; rather the 2nd and 3rd changes are actually \emph{necessary} in order for the 1st change to be consistent.\footnote{
	In \cite{Soni:2024}, we will see that the third change is not relevant away from pure GR. But we do still need to make a change for the inner horizon---it is the even more radical change of replacing $\bar{A}$ with $+\infty$.
}

To see the physical significance of change \#2, consider a trapped surface $\cal T$ lying inside the event horizon.
Such surfaces have $\theta_{(l)} < 0$ and $A({\cal T}) < A_+$, thus there is a risk that they will provide an entropy bound that is too tight.
The Bousso bound is saved by the fact that because $\theta_{(l)} < 0$, by the Penrose singularity theorem \cite{Penrose:1964wq}, a singularity will inevitably form within an affine time proportional to $1/|\theta_{(l)}|$.
Therefore, the lightsheet $N$ is short, limiting the amount of time for entropy to fall across it --- and if you try to make up for this by stuffing a larger flux of matter across $N$, this matter also has a positive energy which causes $N$ to contract even faster!
Hence one can argue that the bound is satisfied for $\theta_{(l)} < 0$, and in fact that one can come closest to violating the bound in the marginal case when $\theta_{(l)} = 0$.

However, in the case of the HCEB the null surface $N$ plays no significant role (change \#1).
Instead the HCEB is salvaged by the fact that it only equals the area for marginally trapped surfaces with $\theta_{(l)} = 0$.
For trapped surfaces, it exceeds the area at leading order.
Thus, the reason that the HCEB is unviolated by trapped surfaces differs significantly from Bousso's original analysis.

Regarding change \#3, we have already seen how this change is needed to avoid violations near the inner horizon.
In Bousso's original analysis, once again this complication would not have been needed; one can accept that $A_-/4G_{N}$ bounds the entropy falling across the inner horizon, since this is a different place than the outer horizon.
In the HCEB this excuse is, once again, unavailable.

One might think that these epicycles result in a bound which is less elegant than the original Bousso construction.
Perhaps this is true, but we should also highlight the other very important distinction between the two bounds, namely
\begin{enumerate}
	\item[4.] The HCEB can be \emph{derived} from the spectrum of the Cauchy slice theory, not conjectured to avoid counterexamples.
\end{enumerate}
From this perspective, the greater complexity of the bound does not count as evidence against it, since it is an \emph{output} of the $T\overline{T}$ model rather than an input.

\subsubsection{Outer Entropy} \label{sssec:outer-ent}
Unlike the Bousso bound, the outer entropy \cite{Engelhardt:2018kcs,Nomura:2018aus,Bousso:2018fou,Bousso:2019dxk,Wang:2020vxc} can be (a) derived and (b) is a bound on the UV entropy rather than the entropy of some matter fields.
It is the maximum fine-grained entropy consistent with data not just at a codimension-two surface $\sigma$ but also its entire outer wedge $O_{W} [\sigma]$.
For a normal or marginal surface with $\theta_{(k)}, - \theta_{(l)} \ge 0$, the outer wedge is the wedge bounded by the congruences generated by the vectors $k$ and $-l$.

The outer entropy is much more closely related to the HCEB than the Bousso bound.
They differ in two ways
\begin{enumerate}
  \item The outer entropy is a bound holding an entire outer wedge fixed, not just a codimension-two surface.
  \item The outer wedge (and therefore the outer entropy) is not defined for (anti)trapped surfaces or supercritical normal surfaces, or even surfaces that are (anti)trapped in a small region.
\end{enumerate}

The outer entropy and HCEB agree for outer-minimising surfaces.
$\sigma$ is outer-minimising if it is the smallest area surface in $O_{W} [\sigma]$, in the same homology class as $\sigma$.
For non-outer-minimising surfaces, the outer entropy is tighter than the HCEB.
This is as it should be, because more data is being held fixed in the outer entropy case.

For (anti)trapped surfaces $\mathcal{T}$, neither wedge spacelike to $\mathcal{T}$ can be called outer.
The same is true for supercritical surfaces, which are not spacelike-separated from the boundary.
In \cite{Soni:2024}, we will find that the HCEB for these surfaces is infinite in the presence of bulk matter.
The entropy bound for an (anti)trapped surface $\mathcal{T}$ holding either of the spacelike wedges fixed will be the area of the minimal marginal surface homologous to $\mathcal{T}$ in that wedge.
This is neither the HCEB, nor is it strictly the outer entropy.

\subsubsection{Dynamical Entropy}
The recently proposed dynamical black hole entropy \cite{Hollands:2024vbe,Visser:2024pwz} is a special case of our formula for GR.
These authors looked for a version of black hole entropy that satisfied the first law away from equilibrium.
The Bekenstein-Hawking entropy satisfies the first law only for quasi-static processes.
They succeeded in finding such a formula at leading order in perturbation theory around a stationary black hole.
We show here that this formula is nothing but the HCEB of the event horizon.

For simplicity, we consider perturbations of a static black hole, which will be dealt with in general dimensions in section \ref{ssec:hawking-mass}.
In the background, the event horizon is also an apparent horizon: any cut $\mathcal{C}$ of the event horizon is a marginal surface.
With the perturbation, $\mathcal{C}$ is no longer marginal, but is infinitesimally close to a marginal surface $\mu_{\mathcal{C}}$.
We choose $\mu_{\mathcal{C}}$ to be obtained from $\mathcal{C}$ by a translation in the null direction as in \cite{Hollands:2024vbe,Visser:2024pwz}.
The dynamical entropy of $\mathcal{C}$, at leading order in the perturbation, is the area of $\mu_{\mathcal{C}}$,
\begin{equation}
	S_{\mathrm{dyn}} (\mathcal{C}) = \frac{A \left( \mu_{\mathcal{C}} \right)}{4 G_{N}} = S_{\mathrm{bd}} (\mu_{\mathcal{C}}).
  \label{eqn:S-dyn}
\end{equation}
The second equality is true without assuming spherical symmetry, because the HCEB-saturating geometry can be constructed explicitly \cite{Engelhardt:2018kcs}.
It is easy to see that $S_{\mathrm{bd}} (\mathcal{C})$ differs from $S_{\mathrm{bd}} (\mu_{\mathcal{C}})$ only at second order in the perturbation.
Suppose the vector field that translates $\mathcal{C}$ to $\mu_{\mathcal{C}}$ is $v$.
Both $v$ and $\grad_{i} S_{\mathrm{bd}}$ vanish in the background, since in a static solution the event horizon is foliated by marginal surfaces and $S_{\mathrm{bd}}$ is constant (we will show this for higher dimensions in section \ref{ssec:hawking-mass}).
Therefore $\grad_{v} S_{\mathrm{bd}} = v \cdot \grad S_{\mathrm{bd}}$ is non-zero only at second order; at first order,
\begin{equation}
  S_{\mathrm{dyn}} (\mathcal{C}) = S_{\mathrm{bd}} (\mathcal{C}).
  \label{eqn:dyn-bd-equiv}
\end{equation}

Note that the HCEB is \emph{not} the correct generalisation of the dynamical entropy beyond leading order.
This is because the defining equation of the dynamical entropy is the non-equilibrium first law: its null derivative along the event horizon must be related to the stress-energy flux across the event horizon.
The null derivative of the HCEB, however, only satisfies this when the event horizon is approximately an apparent horizon, as we will explore in \cite{Soni:2024}, which is only true at leading order.

\section{BTZ as a Matrix Product State} \label{sec:btz-tn}
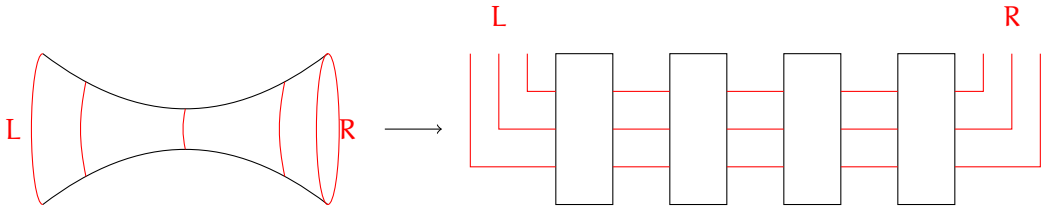
\begin{figure}[h!]
  \centering
	  \begin{tikzpicture}[xscale=.75]
		\coordinate (lt) at (0, 0);
		\coordinate (lb) at (0,-2);
		\coordinate (rt) at (5, 0);
		\coordinate (rb) at (5,-2);
		\coordinate (rc) at (5,-1);

		\draw (lt) to[out=-30,in=-150] coordinate[pos=.15] (1t) coordinate[pos=.5] (2t) coordinate[pos=.85] (3t) (rt);
		\draw (lb) to[out= 30,in= 150] coordinate[pos=.15] (1b) coordinate[pos=.5] (2b) coordinate[pos=.85] (3b) (rb);

		\draw[red] (lb) arc
			[	
			start angle=-90,
			end angle  =-270,
			x radius=.2cm,
			y radius=1cm
			] node[pos=.5,left] {$L$};
		\draw[red] (rc) ellipse (.2 and 1) node[right] {$R$};

		\draw[red] (1t) to[out=-105,in=105] (1b);
		\draw[red] (2t) to[out=-105,in=105] (2b);
		\draw[red] (3t) to[out=-105,in=105] (3b);

		\draw[->] (6,-1) -- (7,-1);

		\begin{scope}[xshift=7cm]
		  \draw (2,-2) rectangle (3,0);
		  \draw (4,-2) rectangle (5,0);
		  \draw (6,-2) rectangle (7,0);
		  \draw (8,-2) rectangle (9,0);

			\begin{scope}[red]
				\node (L) at ( 1,.5) {$L$};
				\node (R) at (10,.5) {$R$};

			  \foreach \y in {-1.5,-1,-.5}
			  {
			    \draw (2,\y) -- (2+\y,\y) -- (2+\y,0);
			    \draw (9,\y) -- (9-\y,\y) -- (9-\y,0);

					\foreach \x in {3,5,7}
					{
					  \draw (\x,\y) -- (\x+1,\y);
					}
			  }
			\end{scope}
		\end{scope}
	\end{tikzpicture}
  \caption{Discretising a Cauchy slice of the BTZ black hole into many segments gives a tensor network interpretation of the bulk. Each codimension-two slice becomes a set of bonds, and each segment of the Cauchy slice becomes a tensor. This tensor network has the same bond dimension throughout.}
  \label{fig:btz-cuts}
\end{figure}

Consider now the case where we cut a Cauchy slice of the BTZ black hole into a large number $N$ of discrete (imaginary) times $\tau_i$ as in figure \ref{fig:btz-cuts}.
We can regulate the left and right timelike CFT boundaries by taking a very small $T\overline{T}$-deformation so that their boundaries are located at a finite distance in the bulk, which we take to be located at $\tau_0$ and $\tau_N$ respectively.

The entangled boundary state dual to the Cauchy slice $\Sigma$ with metric $g$ and extrinsic curvature $K$ is given by the coherent state $\bra{g,T} \hat{P}$ defined in section \ref{ssec:cauchy-bd}.
It is an integral of \eqref{eqn:wdw-state} with some weighting function.
We can rewrite the coherent state $\bra{g,T}$ schematically as
\begin{align}
  \bra{g,T}_\mathrm{bdy} &= \mathds{J}_L \cdot \mathds{T}_{0\to M} [g,T] \cdot \mathds{J}_R \cdot \mathds{1}_{R} \nonumber\\
												 &= \mathds{J}_L \cdot \left( \prod_{i = 0}^{N-1} \mathds{T}_{i \to i+1} \right) \cdot \mathds{J}_R \cdot \mathds{1}_{R}.
\end{align}
where all products represent matrix multiplication, $\mathds{J}_L$ and $\mathds{J}_R$ are the nontrivial junction-cum-scaling operators in \eqref{eqn:junct-map}, and $\mathds{1}_{R}$ is the insertion of the identity in the right boundary Hilbert space $\mathcal{H}_R$, considered as an element of $\mathcal{H}_{R}^{*} \otimes \mathcal{H}_{R}^{*}$.
$\mathds{T}_{0 \to M} [g,T]$ is an integral over the $\mathds{T}[g]$ operator in \eqref{eqn:transition-cft} with a coherent wavefunction.

An important point is that $\mathds{T}_{0 \to M} [g,T]$ is not a local operator.
While $\mathds{T}[g]$ is local (at least in the approximation we are working in), the transformation from it to the $\mathds{T}_{0 \to M} [g,T]$ is not, due to the gravitational constraints.
This contributes to bulk entanglement, as we will discuss more carefully in \cite{Soni:2024}.

Since the holographic transition amplitudes are peaked near some given $\mathcal{E},J$ at each $\tau$, it should also be possible to insert projection operators $P$ between each matrix that requires $\mathcal{E},J$ to lie in some small interval, without changing the state $|g\rangle_\text{bdy}$ by very much.
This effectively replaces each matrix $T_{i \to i+1}$ with a \emph{finite dimensional matrix}, where the dimension at each step is given approximately by the microcanonical entropy
\begin{equation}
  \log \operatorname{dim} \mathcal{H}_{i} \approx S(\mathcal{E}_i, J_{i})
\end{equation}
up to subleading corrections, making this literally a Matrix Product State.

This replacement is not innocent.
In the path integral $\braket{g}{\Psi}$, the bound on entropy at $\sigma \in \Sigma$ comes from the projection on $\bra{g}$ induced by $\ket{\Psi}$, as discussed in section \ref{sec:bound-gen}.
However, after this replacement, the discretised Cauchy slice \emph{itself} has a finite capacity for entropy.
And this is the reason we can call it a tensor network: a tensor network is a representation of the wavefunction of the boundary state, and so the entropy bound cannot come from a projection onto $\ket{\Psi}$.
The tensor network in figure \ref{fig:btz-cuts} \emph{creates} the microcanonically entangled state of the two boundaries.
The passage from a fixed metric state to a coherent state is what allows this --- in other words, we recover the eminently agreeable statement that fixed-metric states are not classical but only coherent states are.

A rather amazing thing about this tensor network, as opposed to previous proposals for holographic tensor networks, is that --- assuming no matter or non-trivial topology --- the bond dimension is constant throughout the network.
This is true even when the Cauchy slice goes behind the horizon of a black hole.
This reflects the fact that 3d GR has no local degrees of freedom, and so there is nothing local to constrict information flow.\footnote{Since the bond dimension is merely an upper bound, there are various ways in which one might modify these tensor networks by \emph{increasing} the bond dimension.  This would amount to including (unnecessary) additional states in the Hilbert spaces, with values of $\bar{E}$ other than the physically correct one.  Some such tensor networks might have $S_{bd} = A/4G_N$ outside the event horizon, but it cannot have this value \emph{inside} the horizon.}

The main advantage of this tensor network, and similar ones that we will build in \cite{Soni:2024}, is that it can discretise \emph{any} Cauchy slice.
The precise form of our bound is quite necessary for this.
A tensor network has the property that the bond dimension of \emph{any} link is greater than the fine-grained entropy.
This is \emph{not} a property shared by geometric area, since the HRT surface is locally \emph{maximal} in a time-like direction.
Thus, the correct object to identify with the bond dimension is a true entropy bound, like the HCEB.

The main disadvantage of this network with respect to others in the literature is that it encodes the entropy flow between a fixed pair of subsystems of the boundary (here, the two CFTs).
This shortcoming is due to the fact that the entropy bound is not a local integral on $\sigma$, as we see in \eqref{eqn:S-bd}.
As discussed in \cite{Akers:2024wab} (following \cite{Bao:2018pvs}), this non-locality is due to the fact that the edge modes of gravity that contribute to the area term are multipartite-entangled.
Ref.~\cite{Akers:2024wab} discusses an example of a tensor network that encodes multipartite edge modes, and one way forward is to look for similar constructions.

When we add bulk matter possessing local degrees of freedom, the story for trapped surfaces changes, making it difficult to identify a finite bond dimension.  This case will be considered in detail in a followup paper \cite{Soni:2024}.

\section{Extensions} \label{sec:extensions}
Finally, sections \ref{ssec:hrt}, \ref{ssec:multi-surface} and \ref{ssec:hawking-mass} discuss various generalisations: to entropies of subregions of a single CFT, to multi-component surfaces and to higher dimensions.
These three generalisations can be `composed:' the considerations below can, for example, be applied to the entropy of a union of subregions of higher-dimensional CFTs.

\subsection{Surfaces Homologous to a Boundary Interval} \label{ssec:hrt}
In previous sections, we focused on closed surfaces, which can be homologous to an entire CFT.
It is not difficult to generalise to surfaces homologous to a single boundary interval $R$.
The main complication is that such surfaces have infinite length.

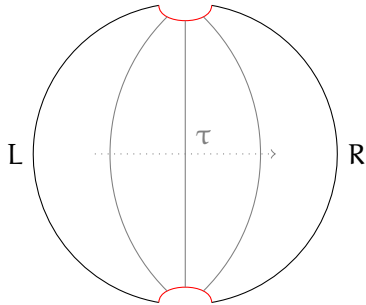
\begin{figure}[h!]
  \centering
		\begin{tikzpicture}
		\coordinate (centre) at (0,0);
		\path (0,0)++(-100:2) coordinate (d1L);
		\path (0,0)++( 100:2) coordinate (d2L);
		\path (0,0)++( -80:2) coordinate (d1R);
		\path (0,0)++(  80:2) coordinate (d2R);

		\draw (d1L) arc (-100:-260:2) node[midway, left] {$L$};
		\draw (d1R) arc ( -80:80:2)   node[midway,right] {$R$};

		\draw[red] (d1L) to[out= 90,in= 90] coordinate[pos=.25] (11) coordinate[pos=.5] (12) coordinate[pos=.75] (13) (d1R);
		\draw[red] (d2L) to[out=-90,in=-90] coordinate[pos=.25] (21) coordinate[pos=.5] (22) coordinate[pos=.75] (23) (d2R);

		\draw[gray,thin] (11) to[out=135,in=-135] coordinate[pos=.5] (tl) (21);
		\draw[gray,thin] (13) to[out= 45,in= -45] coordinate[pos=.5] (tr) (23);
		\draw[gray,thin] (12) to[out= 90,in= -90] (22);

		\path (tl)++(180:.2) coordinate (tll);
		\path (tr)++(  0:.2) coordinate (trr);

		\draw[dotted,gray,->] (tll) -- node[pos=.6,above] {$\tau$} (trr);
	\end{tikzpicture}
  \caption{We excise a small neighbourhood of $\partial A$.}
  \label{fig:regulator}
\end{figure}

To deal with this, we excise a small neighbourhood of $\partial R$ so as to preserve the local hyperbolic symmetry at the asymptotic boundary, as shown in figure \ref{fig:regulator}.
In that case, a surface homologous to $R$ has finite length.
The boundary conditions for the deformed theory on the Cauchy slice can be taken to be the deformed Cardy conditions, defined in \cite{Cardy:2018sdv}.\footnote{Deformed Cardy boundary conditions can behave oddly at finite $\lambda/r^{2}$, but these oddities don't emerge at $\lambda/r^{2} \to 0$ where we are using them. We thank Evan Coleman and Sungyeon Yang for discussions about this.}

In symmetric cases, the induced metric on a Cauchy slice $\Sigma$ can be written as
\begin{equation}
  ds^{2} = \ell^{2} d\tau^{2} + r(\tau)^{2} dx^{2}, \qquad x \in [0,1].
  \label{eqn:rt-cauchy-g}
\end{equation}
It is then easily checked that the formulae in section \ref{ssec:non-const} apply without modification and therefore so do those in section \ref{ssec:entropy}.
The Cardy boundary condition sets $J = 0$.

The (bulk) IR divergence of the area appears as follows.
Since the process of taking away the regulator leaves the extrinsic curvatures in the bulk unaffected, we have to take the limit $r \to \infty$ in \eqref{eqn:tube-T} while keeping $U,V$ finite.
This requires taking $\mathcal{E} \to \infty$ as well, and therefore also the entropy.
This mechanism for the IR divergence was also found in \cite{Lin:2021veu,Wong:2022eiu}.
In the non-symmetric case, we again use an open Wilson line, as in \cite{Ammon:2013hba,Castro:2018srf,Wong:2022eiu}.

\subsection{Multi-Component Surfaces} \label{ssec:multi-surface}
Now, we turn to multi-component surfaces.
These turn up whenever we are interested in calculating the entropy of a multi-component region on the boundary.
Two examples of special interest are multiple intervals, and multi-boundary wormholes.
For ease of discussion, we focus on the latter case; the considerations in the former case are similar except for the fact that we have to deal with IR divergences.

\begin{figure}[h!]
  \centering
		\begin{tikzpicture}[scale=.7]
		\coordinate (c1) at (0, 0);
		\coordinate (c2) at (4, 3);
		\coordinate (c3) at (4,-3);

		\coordinate (c1t) at ($(c1)!2cm!(0, 2)$);
		\coordinate (c1b) at ($(c1)!2cm!(0,-2)$);
		\coordinate (c2t) at ($(c2)!2cm!(2, 5)$);
		\coordinate (c2b) at ($(c2)!2cm!(5, 2)$);
		\coordinate (c3t) at ($(c3)!2cm!(5,-2)$);
		\coordinate (c3b) at ($(c3)!2cm!(3,-4)$);

		\draw[thick] (c1b) arc
			[	
			start angle=-90,
			end angle  =-270,
			x radius=1cm,
			y radius=2cm
			] node[pos=.5,left] {$A$};
		\draw[thick,rotate around={ 45:(c2)}] (c2) ellipse (1 and 2) node[right] {$B$};
		\draw[thick,rotate around={-45:(c3)}] (c3) ellipse (1 and 2) node[right] {$C$};

		\draw[thick] (c1t) to[out=  0,in=-135] coordinate[pos=.2] (at) coordinate[pos=.8] (bt) (c2t);
		\draw[thick] (c1b) to[out=  0,in= 135] coordinate[pos=.2] (ab) coordinate[pos=.8] (cb) (c3b);
		\draw[thick] (c2b) to[out=215,in= 135] coordinate[pos=.2] (bb) coordinate[pos=.8] (ct) (c3t);

		\draw[gray] (at) to[out=-135,in= 135] node[right] {$\gamma_{A}$}(ab);
		\draw[gray] (bt) to[out=-120,in=-120] node[ left] {$\gamma_{B}$}(bb);
		\draw[gray] (ct) to[out= 120,in= 120] node[ left] {$\gamma_{C}$}(cb);
	\end{tikzpicture}
  \caption{A three-boundary wormhole. The HRT surface for $A \cup B$ might have one component ($\gamma_{C}$) or two ($\gamma_{A} \cup \gamma_{B}$).}
  \label{fig:3-bd-worm}
\end{figure}
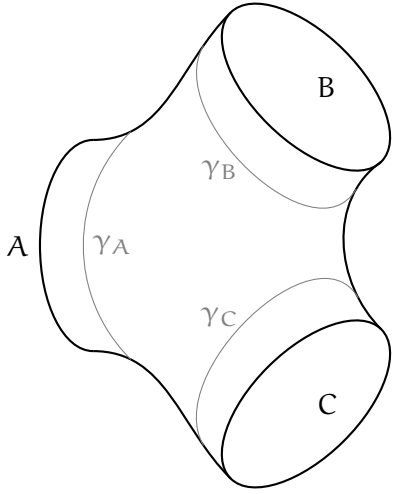

Consider a three-boundary wormhole as shown in figure \ref{fig:3-bd-worm}, and suppose we want to calculate the entropy of $A \cup B$.
In this case, the holographic entropy is known to be:
\begin{equation}
  S_{AB} = \min \left( \frac{2\pi r_{C}}{4 G_{N}}, \frac{2\pi \left( r_{A} + r_{B} \right)}{4 G_{N}} \right),
  \label{eqn:2-bd-ent}
\end{equation}
where $r_{A,B,C}$ are the radii of the three event horizons.
We can attempt to calculate the entropy both on single-component surfaces homotopic to $C$ and on two-component surfaces homotopic to $A \cup B$.
In the former case, we follow the logic of section \ref{sec:ent-bd} and reproduce $2\pi r_{C}/ 4G_{N}$.

The latter case is the one that poses a new challenge.
First, we solve for $\mathcal{E},J$ separately on each component $\sigma_{A}$ and $\sigma_{B}$, where the subscript denotes the boundary region the surface is homotopically related to; suppose we obtain $(\mathcal{E}_{A}, J_{A})$ and $(\mathcal{E}_{B}, J_{B})$.
The way to assign an entropy to $\sigma_{A} \cup \sigma_{B}$ is to go back to the MaxEnt logic from section \ref{ssec:cft-bd}.
It is easy to see that the maximum entropy state consistent with the data is a factorised microcanonical ensemble.
This means that
\begin{equation}
  S_{\mathrm{bd}} (\sigma_{A} \cup \sigma_{B}) = S_{\mathrm{bd}} (\sigma_{A}) + S_{\mathrm{bd}} (\sigma_{B}).
  \label{eqn:two-comp-bd}
\end{equation}
Amazingly, this is exactly consistent with AdS/CFT; the second term in \eqref{eqn:2-bd-ent} is just the sum of the maximum entropy that can flow through the two horizons.

The generalisation to a larger number of components and the case of boundary intervals follows straightforwardly.
The only new rule is that the entropy bound is additive across components.

\subsection{Higher Dimensions: The HCEB from the Hawking Mass} \label{ssec:hawking-mass}
Consider a $d+1$-dimensional spacetime that is a saddle-point of the Einstein-Hilbert action without matter
\begin{equation}
  16 \pi G_{N} S = \int_{\mathcal{M}} \left( R - 2 \Lambda \right) + 2 \int_{\partial \mathcal{M}} K, \qquad \Lambda \equiv - \eta \frac{d (d-1)}{2 \ell^{2}}, \ \eta \in \left\{ -1,0,1 \right\}.
  \label{eqn:eh-S}
\end{equation}
We restrict to solutions foliated by codimension-2 surfaces having maximal symmetry (spherical, flat, or hyperbolic, depending on the sign of the curvature).
Examples where the surfaces have positive curvature are a Schwarzschild black hole and vacuum de Sitter; an example where they have negative curvature is an asymptotically AdS hyperbolic black hole (surfaces homologous to a boundary $S^{d-1}$ in vacuum AdS are included here \cite{Casini:2011kv}).

We can parametrise such a solution as
\begin{equation}
  ds^{2} = - 2 e^{- f(x^{+}, x^{-})} dx^{+} dx^{-} + r\left( x^{+}, x^{-} \right)^{2} d\Omega_{k}^{2},
  \label{eqn:symm-g}
\end{equation}
where $d\Omega_{k}^{2}$ is the metric of a unit radius symmetric space; the space is $S^{d-1}$ for $k=1$, $\mathds{R}^{d-1}$ for $k=0$, and $H^{d-1}$ for $k=-1$.
For $d = 2$, $k$ is always $0$, but the surface can be either $S^{1}$ or $\mathds{R}$.
The vacuum Einstein equations are
\begin{align}
  \frac{r}{d-1} E_{\pm\pm} &= - \partial_{\pm}^{2} r - \partial_{\pm} f \partial_{\pm} r \nonumber\\
  \frac{r}{d-1} E_{+-} &= \partial_{+} \partial_{-} r + \frac{\eta}{\ell^{2}} \frac{d}{2} r e^{-f} + \frac{d-2}{2r} \left( k e^{-f} + 2 \partial_{+} r \partial_{-} r \right).
  \label{eqn:eins-eqn}
\end{align}

We can define a variant of the Hawking mass that is conserved in a presence of the cosmological constant as \cite{Folkestad:2022dse,Folkestad:2023cze}
\begin{align}
	\mu \left( x^{+}, x^{-} \right) &\equiv r^{d} \left[ \frac{k}{r^{2}} + \frac{2 e^{f} \theta_{+} \theta_{-}}{ \left( d-1 \right)^{2}} + \frac{\eta}{\ell^{2}} \right], \qquad \theta_{\pm} = \left( d-1 \right) \frac{\partial_{\pm} r}{r} \nonumber\\
	\partial_{\pm} \mu &= - \frac{16 \pi G_{N}}{d-1} r^{d-1} e^{f} \left( t_{\pm +} \partial_{-} r + t_{\pm -} \partial_{+} r \right) \xrightarrow{t_{ij} = 0} 0.
  \label{eqn:hawking-mass}
\end{align}
Here $\theta_{\pm}$ are the expansions along the $x^{\pm}$ directions.
In $d = 2, \eta=-1$, the case we have dealt with in the bulk of this work,
\begin{equation}
  \mu  = \frac{4\lambda}{\ell^{2}} \mathcal{E} = \frac{16\pi G_{N}}{\ell} \mathcal{E}.
  \label{eqn:hawking-3d}
\end{equation}

At an extremal surface where $\theta_{\pm} = 0$,
\begin{equation}
  \mu  = r_{*}^{d} \left[ \frac{k}{r_{*}^{2}} + \frac{\eta}{\ell^{2}} \right]
  \label{eqn:mu-ext}
\end{equation}
We can calculate the Hawking mass on any codimension-two surface $\sigma$ and use this equation to solve for the radius of the extremal surface.
There will in general be multiple solutions; let us call the maximal one $r_{+}$.
Then the HCEB for a given $\sigma$ is simply
\begin{equation}
  S_{\mathrm{bd}} (\sigma) = \frac{\Omega_{d-1,k} r_{+}(r,\theta_{\pm},k,\eta)^{d-1}}{4 G_{N}},
  \label{eqn:d-bd}
\end{equation}
where $\Omega_{d-1,k}$ is the volume of the unit radius symmetric space with curvature $k$.
The choice of the maximal solution of \eqref{eqn:mu-ext} is the analogue of maximising between the area and the dual area in $d=2$.

\section{Discussion} \label{sec:disc}
In this work, we have proposed a new covariant entropy bound, which we call the holographic covariant entropy bound.
It counts the number of states flowing through a codimension-two surface $\sigma$ in a putative dual theory that lives on a Cauchy slice $\Sigma \supset \sigma$.
Further, this reasoning applies equally well to surfaces inside a black hole.

There are many immediate open questions.
Most importantly, we should understand better the case without maximal symmetry, especially in higher dimensions.
It would be particularly interesting to compare the bound for surfaces that are normal in some places and (anti)trapped in others to the area.
It should also be possible to generalise the techniques of \cite{Akers:2024wab} to construct tensor networks with multiple possible bi-partitions of the boundary.
Finally, it would be interesting to understand better the relation of our story to that of \cite{Bousso:2022hlz,Bousso:2023sya,Bousso:2024ysg}.

\subsection*{Entropy v/s Hilbert Space Dimension}
Some may regard as heretical, our assertion that the area of a surface has \emph{no relation} to the dimension of the Hilbert space that lives on it.
The Hilbert space dimension is always infinite (or, after the sort of UV regulation advocated in \cite{Batra:2024kjl}, a fixed finite number).
Finite notions of entropy arise from knowing further details about correlation functions in the state.
But this property follows naturally if we keep the imaginary eigenvalues in the $T \overline{T}$ spectrum.
And without this modification, it will be impossible to find tensor networks valid behind the horizon.

It is interesting to rephrase this in the spirit of Bekenstein, see \cite{Bousso:2002ju} for a review.
Many such arguments relating entropy to area involve passing matter through $\sigma$ and arguing that a black hole eats up $\sigma$ when entropy exceeds area.
Our perspective is that, since there are states where the black hole has a smooth interior (and these have high entropies), $\sigma$ passing into the black hole is not a problem.
It would be interesting to explore this further.

\subsection*{A New Factorisation Map for Gravity}

\begin{figure}[h!]
  \centering
	\includegraphics{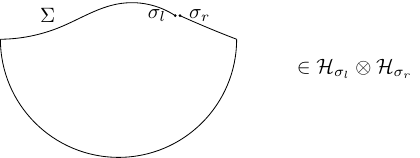}
  \caption{A factorisation map from Cauchy slice holography. Cutting up the path integral that calculates $\braket{g}{\Psi}$ at $\sigma$ creates a state in $\mathcal{H}_{\sigma} \otimes \mathcal{H}_{\sigma}$. This is the factorisation map implicit in the tensor network constructions in the bulk of this paper.}
  \label{fig:fact-map}
\end{figure}

The re-interpretation of the WdW path integral in this work also suggests a new factorisation map \cite{Jafferis:2019wkd,Hung:2019bnq} for gravity \cite{Wall:2021bxi}.
Split the Cauchy slice at $\sigma$ and then consider the WdW wavefunction as a path integral between the two sides of $\sigma$, as in figure \ref{fig:fact-map}.
This factorisation map was the one explored in \cite{Wall:2021bxi}, where it was found (through a bulk version of the arguments in appendix \ref{app:price}) that the Hilbert space obtained by this factorisation map does not have a positive-definite inner product.
It is also related to the factorisation map natural in the BF formulation of JT gravity \cite{Blommaert:2018iqz,Soni:2022}, or the topological field theory model of 3d GR \cite{Mertens:2022ujr,Wong:2022eiu,Akers:2024wab}.
The HCEB is the entanglement entropy in the factorisation map of figure \ref{fig:fact-map}, by construction.

Interestingly, expressions very similar to our formula for the entropy has appeared in \cite{Pulakkat:2024qps} as the Casimir of edge modes of the gravitational phase space.
An important difference of our methods with phase space methods is that we have an explicit Hilbert space; whereas methods based on phase space analysis result in edge mode algebras that could in principle have many representations \cite{Donnelly:2020xgu,Donnelly:2022kfs,Ciambelli:2024qgi}, and this makes it hard to derive an entropy.
It would be interesting to merge these two approaches.

\section*{Acknowledgements}
We thank
Chris Akers,
Goncalo Araujo-Regado,
Raphael Bousso,
Rick Cheng,
Luca Ciambelli,
William Donnelly,
Netta Engelhardt,
Laurent Freidel,
Alex Frenkel,
Ted Jacobson,
Rifath Khan,
Adam Levine,
Juan Maldacena,
Henry Maxfield,
Prahar Mitra,
Onkar Parrikar,
Geoffrey Penington,
Grant Remmen,
Xiao-Liang Qi,
Pratik Rath,
Arvin Shahbazi-Moghaddam,
Vasu Shyam,
Eva Silverstein,
Antony Speranza,
Douglas Stanford,
Ayngaran Thavanesan,
Gonzalo Torroba,
Annie Y. Wei,
Manus R. Visser,
Edward Witten,
Gabriel Wong,
Sungyeon Yang,
and Zihan Yan
for discussions.

This work was partially supported by an AFOSR grant ``Tensor Networks and Holographic Spacetime'', an Isaac Newton Trust Early Career grant ``Quantum Cosmology and Emergent Time'', and STFC consolidated grants ST/T000694/1 and ST/P000681/1.
They also thank UC Berkeley for hospitality.
AW was also partly supported by NSF grant PHY-2207584 while finishing this paper during his sabbatical at the IAS.

\appendix
\section{Notation, Conventions and Standard Formulas} \label{app:convs}
For most of this work, we deal with a ($T \overline{T}$-deformed) two-dimensional holographic CFT that is dual to pure gravity at low energies.
We list some notation here for clarity.

We will use the indices $\mu,\nu$ for indices on a Cauchy slice and $i,j$ for bulk indices.
$\alpha,\beta$ are indices normal to the codimension-two surface of interest $\sigma$.

Turning to 2d CFT, the first piece of notation is for the central charge.
In a 2d holographic CFT, the central charge is related to the Newton's constant of the bulk gravity and the AdS radius $\ell$ as
\begin{equation}
  c = \frac{3 \ell}{2 G_{N}}, \qquad \frac{c}{24\pi} = \frac{\ell}{16 \pi G_{N}}.
  \label{eqn:c-G}
\end{equation}
The second equation here is equivalent, but often easier to remember.

We define the stress tensor of a Euclidean field theory as
\begin{equation}
  \langle \mathcal{T}^{\mu\nu} \rangle = \frac{2}{\sqrt{g}} \frac{\delta}{\delta g_{\mu\nu}} (- \log Z[g]), \qquad \langle \mathcal{T}_{\mu\nu} \rangle = g_{\mu\rho} g_{\nu\sigma} \langle \mathcal{T}^{\rho\sigma} \rangle = - \frac{2}{\sqrt{g}} \frac{\delta}{\delta g^{\mu\nu}} (- \log Z[g]).
  \label{eqn:T-conv}
\end{equation}
Note the often confusing sign difference between the two equations.
The functional derivative with respect to a tensor is defined so that e.g.
\begin{equation}
  \frac{\delta g_{\mu\nu}}{\delta g_{\rho\sigma}} = \tfrac{1}{2}(\delta_\mu^\rho \delta_\nu^\sigma + \delta_\mu^\sigma \delta_\nu^\rho).
\end{equation}
\eqref{eqn:T-conv} is equivalent to the Lorentzian field theory convention\footnote{By equivalent, we mean that $\mathcal{T}^{0}_{0}$ is invariant under Wick rotation.}
\begin{equation}
	\langle \mathcal{T}^{\mu\nu} \rangle = \frac{2}{\sqrt{-g}} \frac{\delta}{\delta g_{\mu\nu}} (i \log Z[g]) \qquad \langle \mathcal{T}_{\mu\nu} \rangle = \frac{2}{\sqrt{-g}} \frac{\delta}{\delta g^{\mu\nu}} \left( -i \log Z[g] \right).
  \label{eqn:T-conv-lor}
\end{equation}
This is opposite to the convention in the GR literature.
With this convention, the Weyl anomaly of a 2d CFT is
\begin{equation}
  \langle \mathcal{T}^{\mu}_{\mu} \rangle = - \frac{c}{24\pi} R
  \label{eqn:anomaly}
\end{equation}
and the energy expectation value of a state on a circle is
\begin{equation}
  E = \langle H \rangle = \int_{S^{1}} \langle \mathcal{T}^{0}_{0} \rangle.
  \label{eqn:energy}
\end{equation}
The relation between the energy and the mixed-index stress tensor is independent of signature.

We will always take the stress tensor of a CFT on a cylinder of metric
\begin{equation}
  \widehat{ds}^{2} = \ell^{2} d \mathsf{t}^{2} + r^{2} d\phi^{2}
  \label{eqn:cyl-g}
\end{equation}
to have the form
\begin{equation}
  \langle \mathcal{T}^{\mu}_{\ \ \nu} \rangle = \left(
    \begin{matrix}
      \frac{\langle \mathcal{E} \rangle}{r^{2}} & i \frac{\langle J \rangle}{\ell r}\\[6pt]
      i \frac{\ell \langle J \rangle}{r^{3}}  & - \frac{\langle \mathcal{E} \rangle}{r^{2}}\\
    \end{matrix}
    \right),
  \label{eqn:cyl-T-cft}
\end{equation}
where $\langle \mathcal{E} \rangle, \langle J \rangle$ are constant.
With the factors of $i$, this is a real angular momentum in Lorentzian space.
This is not quite the most general form, but it \emph{is} the most general form for a superposition over primary states.\footnote{Writing the stress tensor on the plane as $T(z) = \sum_{m \in \mathds{Z}} L_{m} z^{-(m+2)}$, this is the contribution of the $m=0$ term. Angular dependence appears when an $m \neq 0$ term contributes, which only happens when the state is a superposition over at least two distinct states in the same Verma module.}
We can take the state to be a superposition over primaries because in holographic CFTs primaries vastly outnumber descendants at $\mathcal{E} = \mathcal{O} (c)$ \cite{Datta:2019jeo}, which will be the case of interest.
The mixed index stress tensor always has mass dimension $2$; with dimensionless coordinates, the lower-index stress tensor is dimensionless and the upper-index one has mass dimension $4$.
We will find it convenient to use dimensionless coordinates throughout this work.
For an eigenstate, $\mathcal{E},J$ are related to the CFT levels $h, \bar{h}$ as\footnote{The eigenvalue $J$ is an integer, but the expectation value $\langle J \rangle$ may not be. Since we will be interested in $J = \mathcal{O} (c)$, we can ignore this subtlety.}
\begin{equation}
  \mathcal{E} = \frac{h + \bar{h} - \frac{c}{12}}{2\pi}, \qquad J = \frac{h - \bar{h}}{2\pi}.
  \label{eqn:E-J-h}
\end{equation}
The ground state is at $\mathcal{E} = - \tfrac{c}{24\pi}$, and the Hawking-Page level dual to a non-rotating BTZ black hole of radius $r_{h} = \ell$ is $\mathcal{E} = \tfrac{c}{24\pi}$.

The CFT partition function on two Weyl-related metrics on a manifold $\mathcal{M}$ are related by the Liouville action
\begin{equation}
  \log Z [g = e^{2\Omega} \hat{g}] = \log Z[\hat{g}] + \frac{c}{24\pi} \left[ \int_{\mathcal{M}} d^{2} x \sqrt{\hat{g}} \left( R[\hat{g}] \Omega + \hat{g}^{\mu\nu} \partial_{\mu} \Omega \partial_{\nu} \Omega \right) + 2 \int_{\partial \mathcal{M}} dx \sqrt{\hat{\gamma}}\, K[\hat{g}] \Omega  \right].
  \label{eqn:weyl-tr}
\end{equation}
Here, $\hat{\gamma}$ is the determinant of the induced metric on $\partial \mathcal{M}$.
The stress tensor transforms as
\begin{equation}
  \mathcal{T}_{\mu\nu} [g] = \mathcal{T}_{\mu\nu} [\hat{g}] + \frac{c}{12\pi} \left[ \partial_{\mu} \Omega \partial_{\nu} \Omega - \hat{\grad}_{\mu} \partial_{\nu} \Omega + \hat{g}_{\mu\nu} \left( \hat{\Box} \Omega - \frac{1}{2} \hat{g}^{\mu\nu} \partial_{\mu} \Omega \partial_{\nu} \Omega \right) \right].
  \label{eqn:T-weyl}
\end{equation}
This formula becomes the more well-known Schwarzian formula when the Weyl transformation is also a conformal transformation.
Unlike previous formulas, this is an operator equation --- the anomalous conformal symmetry ensures that the stress tensor operator changes by a c-number.

The final CFT formula is the Cardy entropy, which states that for 2d CFTs the density of states at large enough $\mathcal{E}$ takes the form
\begin{equation}
  \log \rho(\mathcal{E}) = 2\pi \sqrt{\frac{\pi c}{3} \left( \mathcal{E} + \sqrt{\mathcal{E}^{2} - J^{2}} \right)}.
  \label{eqn:cardy-dos}
\end{equation}
For most CFTs, this is valid when $\mathcal{E} \gg c$, but for holographic CFTs there is an extended range of validity --- due to the sparseness of the low-energy spectrum --- and this holds for $\mathcal{E} \ge \tfrac{c}{24\pi}$ (i.e. the Hawking-Page level) \cite{Hartman:2014oaa,Mukhametzhanov:2019pzy}.

\section{More on the Spectrum} \label{app:spectrum}

\subsection{The Price of Self-Adjointness}\label{app:price}

One might think that since the spectrum contains states with $\text{Im}(E) \ne 0$, the Hamiltonian $H$ of the theory is not self-adjoint.
But there is a mathematically nicer alternative, which is to say that some of the states in the model, including the energy eigenvectors for which $\text{Re}(E) > 0$, actually have negative norm.

Recall that the adjoint of an operator is defined as $({\cal O}^{B'}_{A'})^\dagger = I_{A,{A'}} I^{B,{B'}}{\cal O}_B^A$ where $I$ is the inner product or its inverse.
Hence, on a Krein space ${\cal K}$ (which is similar to a Hilbert space but it allows negative norm states) the definition of the adjoint differs; and in particular it allows for a self-adjoint Hamiltonian $H = H^\dagger$ to have complex energy eigenvalues, as long as (i) the eigenvectors are null vectors with respect to $I$, and (ii) the eigenvalues come in complex conjugate pairs with energy $E$ and $E^*$ respectively, with a nonzero inner product between these vectors.

For a simple toy model illustrating the relevant features, consider the Krein space ${\cal K}_{1,1}$ which in an orthonormal basis has the following inner product:
\begin{equation}
	I_{A,A'} = \left(\!
		\begin{tabular}{cccc} 
			$1$ & $0$ \\
			$0$ & $-1$
		\end{tabular}
	\!\right).
\end{equation}
Then the most general traceless self-adjoint Hamiltonian is
\begin{equation}\label{toyH}
	H = \left(\!
		\begin{tabular}{cccc} 
			$a$ & $b$ \\
			$-b^*$ & $-a$
		\end{tabular}
	\!\right)
\end{equation}
with $a \in \mathds{R}$, $b \in \mathds{C}$.
Note the relative sign of the off-diagonal terms, which is the opposite from what you expect.
We have chosen for $H$ to be traceless to replicate the $E \to -E$ symmetry of the $T\overline{T}$ spectrum.
The energy eigenvalues are then $(E, -E)$ with
\begin{equation}
	E = \pm \sqrt{-\mathrm{det} H} = \sqrt{a^2 - b^{*}b}
\end{equation}
Thus, instead of the usual eigenvalue repulsion, there is an eigenvalue \emph{attraction} between energy eigenvectors with opposite norm, leading to imaginary eigenvalues when $|b| > |a|$.
At $|b| = |a|$ there is a branch point, and the eigenvalues are $(0,\eta)$ where $\eta$ is Grassmanian.
Yet despite the non-smoothness of the $E$-values, the components of $H$ as an operator are completely smooth through the transition.

This suggests that if we want to find a microscopic theory in the UV which gives rise to the $T\overline{T}$ deformation as an effective field theory, we should generalize QFT to the case of an indefinite norm.

Since the original CFT states all have positive norm, and since eigenvalue attraction only works between states with opposite norm, it is evident that states with $\text{Re}(E) > 0$ must be assigned a negative norm.
It thus appears that the $E \to -E$ symmetry of the spectrum also reverses the sign of the inner product $I_{A,A'}$.
Oddly this means that $\langle E \rangle_\psi \le 0$ for all states $\psi$ on the real axis.

Normally in QM a symmetry must be implemented with some specific operator.
In our ${\cal K}_{1,1}$ model, without loss of generality we can take $E$-reversal symmetry to be
\begin{equation}\label{Erev}
	R = \left(\!
		\begin{tabular}{cccc} 
			$0$ & $1$ \\
			$1$ & $0$
		\end{tabular}
	\!\right),
\end{equation}
which then fixes $b$ to be real so that $R H R = -H$.
Note that $R = -R^\dagger$, so formally $RHR^\dagger = H$, just as if it were a unitary symmetry of the Hamiltonian.\footnote{
	What if we instead use
	\begin{equation}\label{Erev}
		R = \left(\!
			\begin{tabular}{cccc} 
				$0$ & $i$ \\
				$i$ & $0$
			\end{tabular}
		\!\right)?
	\end{equation}
	Then $R^2 = -1$, $R = R^\dagger$, but $RHR^\dagger = H$, $HR = -RH$ just as before.	$e^{iR} H e^{-iR}$ is now a symmetry of $H$.
}

\subsection{Symmetry Group and Continuity Across the Cut}\label{sapp:cont}

Let us now consider a situation where, by continuously adjusting $\lambda$ or $L$, eigenvalues cross over the branch point.
Can we consistently identify the states before and after the transition?
\eqref{cross} by itself offers no guidance, since there is a degeneracy at $E = 0$.

Fortunately, when the CFT spectrum $({\cal E}, J)$ is non-degenerate, symmetry comes to the rescue.
The key point is that by nondegeneracy we only need to worry about 2 eigenvalues entering the branch point at once.
Whenever this happens, the $\mathcal{K}_{1,1}$ toy model in appendix \ref{app:price} literally describes the situation, since \eqref{toyH} is the most general 2-state Hamiltonian with the relevant symmetries.

Since $J$ is conserved, let us focus on discrete symmetries which preserve $J$ and act nontrivially on $E$.
This should be a 4-element discrete symmetry group $\mathds{Z}_2 \times \mathds{Z}_2$:
\begin{eqnarray}
	CPT: &E \to E^*\:\:,\:\: i \to -i& \\
	R:& E \to -E,\:\:I \to -I&
\end{eqnarray}
Neither symmetry preserves the inner product, since one is antiunitary while the other is norm-reversing.

If we think of the Krein space $\mathcal{K}_{1,1}$ as having 4 \emph{real} dimensions, we can decompose it into 4 distinct irreducible representations of $\mathds{Z}_2 \times \mathds{Z}_2$.
For example, using the same basis as in the previous section, the irrep vectors are
\begin{equation}
	V_+^+ = \left(\!
		\begin{tabular}{cccc} 
			$1$ \\
			$1$  
		\end{tabular}
	\!\right),\:\: 
	V_+^-
	= \left(\!
		\begin{tabular}{cccc} 
			$i$ \\
			$i$  
		\end{tabular}
	\!\right),\:\:
	V_-^+
	= \left(\!
		\begin{tabular}{cccc} 
			$1$ \\
			$-1$  
		\end{tabular}
	\!\right),\:\:
	V_-^-
	= \left(\!
		\begin{tabular}{cccc} 
			$i$ \\
			$-i$  
		\end{tabular}
	\!\right),
\end{equation}
where the top index gives the sign change under $CPT$ and the bottom index gives the sign change under $R$.
All four of these states are null, due to the fact that they involve equal-sized superpositions of positive- and negative-energy states.

When the $\mathcal{K}_{1,1}$ states move from the real to the imaginary axis, symmetry guarantees that we can match each irrep to the corresponding one on the other side, and furthermore the norm must be preserved.
This fixes the relation between the two sides up to a single real parameter $\alpha$,
\begin{align}
	W_+^+ = \alpha\,V_+^+, \qquad&\qquad  W_+^- = \alpha\,V_+^-,\nonumber\\
	W_-^+ = \frac{V_{-}^{+}}{\alpha}, \qquad &\qquad  W_-^- = \frac{V_{-}^{-}}{\alpha},
\end{align}
corresponding to a boost which mixes the positive and negative energy states.
This boost would be generated by the Hamiltonian
\begin{equation}\label{boost}
	H' = \left(\!
		\begin{tabular}{cccc} 
			$0$ & $i$ \\
			$i$ & $0$
		\end{tabular}
	\!\right)
\end{equation}
which is not symmetric under $RH'R = -H'$ and therefore cannot arise as a term in the dynamics.

More generally for any $E$-reversal symmetric dynamics (even with more states), we have $\{R,H\} = 0$ and therefore the Hamiltonian is invariant under an additional boost symmetry:
\begin{equation}
	H = e^{R} H e^{-R}
\end{equation}
Since this is a symmetry of the Hamiltonian, the choice of $\alpha$ is arbitrary, and we can simply choose to normalize our $W$ states so that $\alpha = 1$ across the branch point.

\section{A Second Argument for the Equality of the Bounds} \label{app:eq-2}
In this appendix, we give a second argument that the two questions in sections \ref{ssec:cauchy-bd} and \ref{ssec:bulk-bd} give the same bound.
Unlike the argument in section \ref{ssec:cauchy-bd}, which only uses the nature of coherent states, this argument uses end-of-the-world branes and is therefore less universal.
This argument does however illustrates the shift of perspective that Cauchy slice holography requires, and may be interesting to readers for that reason.
We again assume that the theory is pure 3d GR, but we expect that the argument will generalise more-or-less straightforwardly.

Since the inner product $\braket{g}{\Psi}$ is a path integral on a manifold without boundary, we can re-interpret it as the trace of an operator in the Hilbert space of a single CFT,
\begin{equation}
  \braket{g}{\Psi} = \tr_{R} \hat{T} [g,\Psi] = \sum_{J} \int d \mathcal{E}\, e^{S (\mathcal{E},J)} \mel{\mathcal{E},J}{\hat{T}[g,\Psi]}{\mathcal{E},J}.
  \label{eqn:wdw-new-arrow}
\end{equation}
The second equality here contains an assumption that the matrix element is the same for all states in a microcanonical energy window.

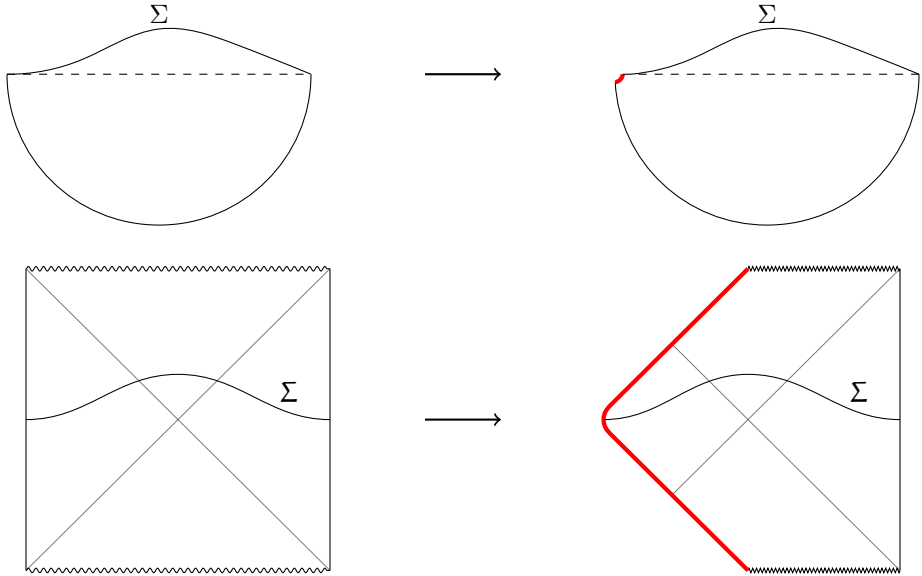
\begin{figure}[h!]
  \centering
	\includegraphics{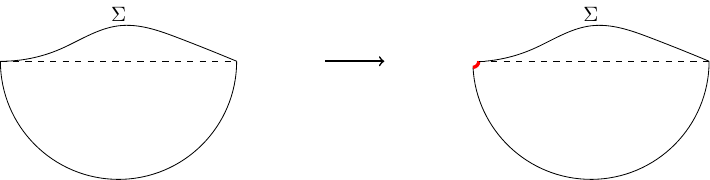}

	\vspace{.5cm}

	  \begin{tikzpicture}
		\begin{scope}[scale=2,xshift=-.5cm]
			\draw (0,0) -- (0,2);
			\draw (2,0) -- (2,2);
			\draw[opacity=.5] (0,0) -- (2,2);
			\draw[opacity=.5] (2,0) -- (0,2);
			\draw[decorate,decoration={snake,amplitude=.3mm,segment length=1mm}] (0,2) -- (2,2);
			\draw[decorate,decoration={snake,amplitude=.3mm,segment length=1mm}] (0,0) -- (2,0);
			\draw[thin] (0,1) to[out=0,in=180] (1,1.3) to[out=0,in=180] node[pos=.75,above] {$\Sigma$} (2,1);
		\end{scope}

		\draw[->,thick] (4.25,2) -- (5.25,2);

		\begin{scope}[xshift=6.5cm]
			\draw (4,0) -- (4,4);
			\draw[opacity=.5] (1,1) -- (4,4);
			\draw[opacity=.5] (1,3) -- (4,0);
			\draw[decorate,decoration={zigzag,amplitude=.3mm,segment length=.5mm}] (2,0) -- (4,0);
			\draw[decorate,decoration={zigzag,amplitude=.3mm,segment length=.5mm}] (2,4) -- (4,4);
			\draw[red,ultra thick,name path=etw] (2,0) -- (.2,1.8) to[out=135,in=-90] (0.1,2) to[out=90,in=-135] (.2,2.2) -- (2,4);
			\draw[thin] (0.1,2) to[out=0,in=180] (2,2.6) to[out=0,in=180] node[pos=.75,above] {$\Sigma$} (4,2);
		\end{scope}

	\end{tikzpicture}
  \caption{We replace the original two-boundary spacetime with a new one-boundary spacetime, with an end-of-the-world brane that approaches the left boundary. The new spacetime has a Cauchy slice with effectively the same metric as one in the original spacetime.}
  \label{fig:etw-1}
\end{figure}

How do we argue for this?
We need a way to isolate the matrix element $\mel{\mathcal{E},J}{\hat{T}[g,\Psi]}{\mathcal{E},J}$ from the sum over a microcanonical window, while retaining semi-classical control.
We use one of a class of projectors that is expected to be dual to end-of-the-world branes \cite{Takayanagi:2011zk,Kourkoulou:2017zaj} in the IR; we call the projector $P_{\mathrm{ETW}}$.
Consider now, instead of the bipartite state $\ket{\Psi}$, the one-party state $P_{\mathrm{ETW}} e^{- H_{L} \epsilon} \ket{\Psi}$.
The bulk dual of this state contains an ETW brane that approaches the (absent) left boundary at one point, as in figure \ref{fig:etw-1}.
There is a Cauchy slice whose metric is $g$, except near the boundary where it hits the ETW brane.

In both cases, and independent of the details of the ETW brane,\footnote{
	Details of the brane enter in $\mathcal{O}(\epsilon)$ corrections.
}
the Wheeler-DeWitt wavefunction is calculated by the same geometry in both cases.
Since the different choice of projector correspond to different ETW branes and the projectors are expected to span an over-complete basis of the Hilbert space \cite{Kourkoulou:2017zaj}, we conclude that the matrix elements of $\hat{T}$ are the same for any state in the microcanonical window.\footnote{
	The $e^{S}$ in this case is only a normalisation factor; the Wheeler-DeWitt wavefunction becomes $\mel{\mathcal{E},J}{\hat{T}}{\mathcal{E},J}$ after correct normalisation.
}

\begin{figure}[h!]
	\centering
	\includegraphics{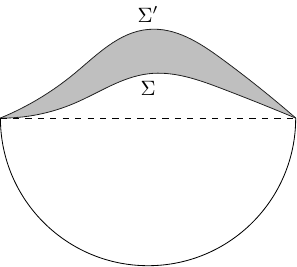}
	\caption{The Wheeler-DeWitt wavefunction for the metrics of two Cauchy slices $\Sigma,\Sigma'$ differs by the on-shell action of the region between the two Cauchy slices.}
	\label{fig:two-cauchy}
\end{figure}

Next, note that if $\sigma$ is extremal \eqref{eqn:S-sigma} is equal to the area.
This means that our entropy bound agrees with the fine-grained entropy at the HRT surface and thus that any slice $\Sigma_{\mathrm{HRT}}$ containing the HRT surface does not project out any of the entanglement in $\ket{\Psi}$.

Finally, we argue that no other slice in the same Lorentzian bulk dual projects out any of the entanglement.
Consider two Cauchy slices $\Sigma,\Sigma'$ with metrics $g,g'$ in the spacetime dual to $\ket{\Psi}$, as illustrated in figure \ref{fig:two-cauchy}.
Then,
\begin{equation}
  \Psi[g'] = e^{i I_{\mathrm{on-shell}} [g,g']} \Psi[g], \qquad \left| \Psi[g] \right|^{2} = \left| \Psi[g'] \right|^{2}.
  \label{eqn:wdw-diff}
\end{equation}
where $I_{\mathrm{on-shell}}$ is the on-shell action of the portion of the spacetime between $\Sigma$ and $\Sigma'$.
This continues to be true when we introduce an ETW brane at the left boundary as in figure \ref{fig:etw-1}, meaning that
\begin{equation}
  \mel{\mathcal{E}_{*},J_{*}}{\hat{T}[g,\Psi]}{\mathcal{E}_{*},J_{*}} = \mel{\mathcal{E}_{*},J_{*}}{\hat{T}[g',\Psi]}{\mathcal{E}_{*},J_{*}},
  \label{eqn:wdw-diff-mel}
\end{equation}
where $\mathcal{E}_{*}, J_{*}$ are related to the expectation value of $T$ in $\ket{\Psi}$.
Thus,
\begin{equation}
	\abs{\frac{\Psi [g]}{\mel{\mathcal{E}_{*},J_{*}}{\hat{T}[g,\Psi]}{\mathcal{E}_{*},J_{*}}}} = \abs{\frac{\Psi [g']}{\mel{\mathcal{E}_{*},J_{*}}{\hat{T}[g',\Psi]}{\mathcal{E}_{*},J_{*}}}} = e^{S(\mathcal{E}_{*},J_{*})}.
  \label{eqn:wdw-diff-S}
\end{equation}
In other words, the entropic term is also the same for $g,g'$ and our argument is complete.

\bibliographystyle{JHEP}
\bibliography{refs.bib}
\end{document}